\documentclass[english]{iopart}

\usepackage{graphicx}
\newcounter{fig}
\newenvironment
{image}[5]
{
\begin{center} 
\newcommand{\fichier}{#1} 
\newcommand{\ang}{#2} 
\newcommand{\zoomera}{#3}
\newcommand{\zoomerb}{#4}
\newcommand{\titre}{#5} 
{
\includegraphics[angle=\ang,width=\zoomera\textwidth,
totalheight=\zoomerb\textwidth]{\fichier} \ \\
\ \\
\stepcounter{fig} F\begin{footnotesize}ig\end{footnotesize}. 
\begin{small}\arabic{fig} $ 
\ $\end{small} \titre
} 
\end{center}
}

\begin{document}

\title[Landau singularities of holonomic integrals of the Ising 
class]{\Large
Landau singularities and
singularities of holonomic integrals of the Ising class}

\author{ 
S. Boukraa$^\dag$, S. Hassani$^\S$,
J.-M. Maillard$^\ddag$ and N. Zenine$^\S$}
\address{\dag LPTHIRM and D\'epartement d'A{\'e}ronautique,
 Universit\'e de Blida, Blida, Algeria}
\address{\S  Centre de Recherche Nucl\'eaire d'Alger, \\
2 Bd. Frantz Fanon, BP 399, 16000 Alger, Algeria}
\address{\ddag\ LPTMC, Universit\'e de Paris 6, Tour 24,
 4\`eme \'etage, case 121, \\
 4 Place Jussieu, 75252 Paris Cedex 05, France} 
\ead{maillard@lptmc.jussieu.fr, maillard@lptl.jussieu.fr, 
 njzenine@yahoo.com, boukraa@mail.univ-blida.dz}

\begin{abstract}

We consider families of multiple and simple integrals of the ``Ising class''
and the linear ordinary differential equations with polynomial
 coefficients they are 
solutions of.
We compare the full set of singularities given by the roots 
of the head polynomial of these linear ODE's and 
the subset of singularities 
occurring in the integrals, with the singularities obtained 
from the Landau conditions.
For these Ising class integrals, we show that the Landau conditions
can be worked out, either to give 
the singularities of the corresponding  linear differential equation
or the singularities occurring in the integral.
The singular behavior of these integrals is obtained in the
self-dual variable $w=\, s/2/(1+s^2)$, with $s=\, \sinh(2K)$,
where $K=J/kT$ is the usual Ising model coupling constant.
Switching to the variable $s$,
we show that the singularities
of  the analytic continuation of series expansions of these 
integrals actually break the Kramers-Wannier duality.
We revisit the singular behavior
(J. Phys. A {\bf 38} (2005) 9439-9474) of the third contribution
to the magnetic susceptibility of Ising model $\chi^{(3)}$ at the points
$1+3w+4w^2=\, 0$ and show that $\chi^{(3)}(s)$ is not singular
 at the corresponding points
inside the unit circle $\vert s \vert=1$, while its
analytical continuation in the variable $s$ is actually singular
at the corresponding points $\, 2+s+s^2=0$ oustside  the unit 
circle ($\vert s \vert > 1$).
\end{abstract} 
\vskip .5cm


\noindent {\bf AMS Classification scheme numbers}: 34M55, 
47E05, 81Qxx, 32G34, 34Lxx, 34Mxx, 14Kxx


\section{Introduction}
\label{intro}

Understanding the magnetic susceptibility $\chi$ of the Ising model as a 
function
of the temperature  remains an outstanding problem in lattice
statistical
physics.
One of the problems raised in this model is the possibility of
occurrence for $\chi$ of a natural boundary  in the complex plane
of the variable $\,s\,=\,\sinh(2K)$, where $\,K=\,J/kT$ is the 
conventional
Ising model coupling constant.
There is a strong argument on this point, since it has been shown by
Nickel~\cite{nickel-99,nickel-00} that the singularities of each
$n-$particle contribution $\, \chi^{(n)}$ to the
susceptibility are given by
\begin{eqnarray}
\label{Nick}
&& s+ {1 \over s}  \,= \, \, \, u^k+u^{-k} \,  \,+ u^m + u^{-m}, \\
&& \hbox{with:} \quad\quad \quad \quad 
 u^n \,= \,1, \quad \quad \quad -n \le k, \, \quad  \, m \le n  \nonumber 
\end{eqnarray}
which are actually lying on the unit circle  $\vert s \vert=1$.
In the following , we will call these singularities
``nickellian singularities''.
These individual $\chi^{(n)}$ are $n-$particle contributions 
to the infinite sum~\cite{wu-mc-tr-ba-76}
\begin{eqnarray}
\chi \,  =  \,\,\sum \, \chi^{(n)} 
\end{eqnarray}
where the singularities (\ref{Nick}) accumulate and (in absence of 
cancellation) densify the unit circle, possibly creating
 a natural boundary for 
$\,\chi$. The above sum 
is restricted to $n$ even (respectively odd)
for the low (respectively high) temperature.

Among these $n-$particle contributions, $\chi^{(1)}$ and $\chi^{(2)}$ 
were known long ago, and the following two, $\, \chi^{(3)}$ 
and $\, \chi^{(4)}$,
have been shown recently~\cite{ze-bo-ha-ma-04,ze-bo-ha-ma-05,ze-bo-ha-ma-05b},
to satisfy Fuchsian linear differential equations
of, respectively, order seven and order ten.

Our study is based on 
a {\em holonomic} approach which uses the fact that, 
the integrand being an {\em algebraic function} of
all the variables, the multiple integral under consideration, can be 
shown to be {\em holonomic}~\cite{Kawai}, i.e.
it satisfies finite order linear differential equations,
with polynomial coefficients, in the remaining
 variable for which no integration has been performed.
The analysis, then, focuses on the corresponding
 linear differential operators.
This holonomic approach gives the full set of singularities as 
roots of the head polynomial in front of the highest derivative of the
differential equation (excluding the "apparent" 
polynomial~\cite{ze-bo-ha-ma-04,ze-bo-ha-ma-05,ze-bo-ha-ma-05b}).

The linear ODE corresponding to  $\chi^{(4)}$ has~\cite{ze-bo-ha-ma-05b},
 besides the known singularities
 $\, s^2 =1$, and $\, s^2 =-1$, only the
 singularities (\ref{Nick}) predicted
by Nickel~\cite{nickel-99, nickel-00}, while the linear ODE
 corresponding to $\, \chi^{(3)}$ has shown
a pair of singularities $\, 1+3w+4w^2=0$ (where $w= \,s/2/(1+s^2)$) 
that are {\em not} on the unit circle $\vert s \vert =1$. 
The condition  $\vert s \vert =1$ is equivalent, for the variable $\, w$,
to be real and such that its absolute value is greater than $\, 1/4$.
In this paper, by $w$ is "in ${\cal W}_c$", we will mean 
$w \in ] -\infty, -1/4] \cup [1/4, \infty[$. 

Therefore, the following question is in order. 
Should we expect  in the linear ODE's corresponding to 
the successive $n$-particle contribution
$\, \chi^{(n)}$, the occurrence of singularities that are of similar
nature as $\, 1\, +3 w\, +4 w^2\, =0$, i.e. not lying in ${\cal W}_c$, 
or only the occurrence of
the "nickellian singularities" given by (\ref{Nick})?

These expected singularities of the linear
ODE governing the $\chi^{(n)}$ are not necessarily
singularities of the particular solution $\chi^{(n)}$. Do 
singularities with $\vert s \vert > 1$ occur in the $\chi^{(n)}$,
and accumulate,  the question of a natural
boundary for $\, \chi$ becomes of the most importance.

In previous works, we have introduced a ``connection 
matrix method''~\cite{ze-bo-ha-ma-05c}  to
determine if a singularity of the linear ODE is actually 
a singularity occurring in the multiple integral.
For the three-particle contribution $\, \chi^{(3)}$, we have
shown~\cite{ze-bo-ha-ma-05c} that the two quadratic roots of
$ \,1+3w+4w^2 \,=\,0$, 
singularities
of the linear ODE, are actually {\em absent in the multiple 
integral defining} $\, \chi^{(3)}$ written as a function
of the self-dual variable $w$\footnote{Be aware
that some subtlety  can be encountered when the variable
$s$ is used, see Section (3.3) and Section (4.5).}.

In our holonomic approach, the singularities of the multiple integrals
are, thus, obtained in two steps: find the linear differential equation, 
then use our ``connection method''~\cite{ze-bo-ha-ma-05c} 
in order to analyze the singularities
of the solutions of this ODE and, especially, the
 singularities of that particular 
solution corresponding to the multiple integral.
These computations may be rather cumbersome,
depending on the order of the linear ODE, on the number of singularities,
and their distribution in the complex plane of the 
variable.

In particle physics, in  the study of analytical properties of the 
$\, S$-matrix \cite{Smatrix}, or in Feynman diagram calculations,
the amplitudes of the Green functions are defined through an integral
representation and depend on external kinematical invariants
(the Mandelstam variables), the quadri-momentum
in the loops being the integration variables.
The singularities of these integrals are associated with physical
thresholds in each of the physical regions in which they describe the 
process under consideration.
The question of forecasting the location of the
singularities of functions defined through integral representations
is, thus, a general and interesting 
problem in mathematical physics~\cite{Kreimer}.

An {\em analytical approach}~\cite{Smatrix} of this
issue provides, by imposing conditions on the integrand, 
singled-out sets of complex points, as {\em possible singularities}
of the {\em integral}.
These conditions on the integrand are called {\em Landau  
conditions}~\cite{Smatrix, landau-59, bjor-drel-65}.
The singularities obtained this way are called 
{\em Landau singularities}.
Note that, the Landau singularities were introduced to be those 
of the {\em integral}, and that, the Landau conditions are only
{\em necessary conditions} for the existence of such
singularities~\cite{Smatrix, bjor-drel-65, nickel-05}. In fact, 
when this analytical $\, S$-matrix approach was flourishing in 
the litterature, it was not yet known~\cite{Kawai}
 that the integrals of (physical) interest 
{\em were actually holonomic},
that is, solutions of finite order linear differential equations with
 polynomial coefficients. The fact that the Landau singularities could 
have a better overlap with the singularities of the linear ODE's,
rather than  with the singularities of the integral of interest,
 was a question nobody 
could imagine at that time. 

In many domains of mathematical physics~\cite{Crandall},
the computation of multiple integrals 
like the evaluation of Feynman diagrams~\cite{Kreimer},
or various correlation functions in quantum field theory
and statistical mechanics~\cite{Boos2006,Korepin}
is an important problem. The $n$-particle contribution
 to the magnetic susceptibility 
of the Ising model have integral 
representations. They are given by $(n-1)$-dimensional 
integrals~\cite{nickel-00,pal-tra-81,yamada-84} that read (omitting the 
prefactor)
\begin{eqnarray}
\label{chi3tild}
\tilde{\chi}^{(n)}\,=\,\,\, {\frac{1}{n!}}  \cdot 
\Bigl( \prod_{j=1}^{n-1}\int_0^{2\pi} {\frac{d\phi_j}{2\pi}} \Bigr)  
\Bigl( \prod_{j=1}^{n} y_j \Bigr)  \cdot   R^{(n)} \cdot
\,\, \Bigl( G^{(n)} \Bigr)^2, 
\end{eqnarray}
where
\begin{eqnarray}
\label{Gn}
G^{(n)}\,=\,\,\,\,\, 
\Bigl( \prod_{j=1}^{n}x_j \Bigr)^{(n-1)/2} \cdot \prod_{1\le i < j \le n}
{\frac{2\sin{((\phi_i-\phi_j)/2)}}{1-x_ix_j}}, 
\end{eqnarray}
and
\begin{eqnarray}
\label{Rn}
R^{(n)} \, = \,\,\, \,  {\frac{1\,+\prod_{i=1}^{n}\, 
x_i}{1\,-\prod_{i=1}^{n}\, x_i}}, 
\end{eqnarray}
with\footnote[5]{Variables $x_i$ and $y_i$ are the variables noted  
$\tilde{x}_i$,
and $\tilde{y}_i$, in previous
 papers~\cite{ze-bo-ha-ma-04,ze-bo-ha-ma-05,ze-bo-ha-ma-05b}.}
\begin{eqnarray}
\label{thex}
&&x_{i}\, =\,\,\,  \,  \frac{2w}{1-2w\cos (\phi _{i})\, 
+\sqrt{\left( 1-2w\cos (\phi_{i})\right)^{2}-4w^{2}}},  
  \\
\label{they}
&&y_{i} \, = \, \, \,
\frac{2w}{\sqrt{\left(1\, -2 w\cos (\phi _{i})\right)^{2}\, -4w^{2}}}, 
\quad
 \quad  \quad  \quad \sum_{j=1}^n \phi_j=\, 0  
\end{eqnarray}
valid for small $w$ and, elsewhere, by analytical continuation.
Recall that small $w$ corresponds to small values of $s$,
 {\em as well as} large values of $s$.

If successful, the Landau conditions will give the singularities of 
these integrals bypassing the linear ODE search and the
singularity analyzis from our connection 
method~\cite{ze-bo-ha-ma-05c}, possibly
providing a deeper insight on the natural boundary problem of the 
susceptibility.

Unfortunatly, the multiple integrals (\ref{chi3tild}) are too involved 
for the Landau conditions to be worked out. We will consider, instead, 
and as a preliminary project, two kinds of functions
in the form of integral representations (one-dimensional 
and multidimensional) which belong to the Ising
class\footnote[8]{The terminology {\em integral of the Ising class} has also
been used by Bailey, Borwein and Crandall in \cite{Crandall}.}.
The integrands of this class of integrals have a 
trigonometric-hyperbolic
form and are simple algebraic functions in the cosinus of the angular
integration variables, and are also algebraic in the
 remaining $\, w$ variable. These integrals
 are thus holonomic: they must satisfy
finite order linear ODE's with polynomial coefficients in
 the remaining variable $\, w$.
The route we take for obtaining the linear ODE's of the two 
sets of integrals will
be through series 
expansions~\cite{ze-bo-ha-ma-04,ze-bo-ha-ma-05,ze-bo-ha-ma-05b}. 

One example corresponds to  
a multiple integral, but yields quite simple sets of singularities,
whereas the second example corresponds to 
simple integrals, but yields quite non trivial
sets of singularities.
These two  ``Ising class'' examples have been chosen  to provide
an ``equivalent level of complexity'' but complementary results on our
singularity problem.
For each example, we compare the singularities of the linear ODE, 
the singularities  of the 
integral, and  the singularities given by the Landau conditions.

The paper is organized as follows.
In Section (\ref{Landau}),
we recall the Landau equations method \cite{Smatrix, itz-zub-80}.
Section (\ref{toy2}) presents the first ``toy model'' for which we can
obtain for {\em any value} $\, n$ of this $\, n$-multiple integral, the
corresponding linear ODE, remarkably of second order,
and the corresponding solutions.
Section (\ref{Diag}) introduces a second ``toy model''
of simple integrals, 
gives the corresponding linear ODE's with their singularities,
presents the Landau singularities, and compares these two sets with
the actual singularities of the integral.
In both sections, we consider some subtleties that appear when we switch
from the self-dual variable $w$ to the
variable $s$, with respect to the analytic continuations of each.
We finally comment the calculations on Landau conditions, sketch
 the forthcoming ``Ising class''
integral calculations we will study, and conclude.

\section{Landau conditions}
\label{Landau}

For the paper to be self contained, we recall, in
this section, general features about the Landau singularities.
Let us, for instance, consider the following integral:
\begin{eqnarray}
\label{theintegral}
F(w) \, =\, \,  \int_D d\phi_1 \cdots d\phi_n \cdot
 f(w, \phi_1, \cdots, \phi_n)
\end{eqnarray}
where $\, f(w, \phi_1, \cdots, \phi_n)$ denotes an algebraic function of $\, w$
and of trigonometric functions of the angles $\,\phi_i$. Denote by
 $B_j \left( w, \phi  \right)$, $j=1,2, \cdots d$,
the set representing the boundary of the integration domain, and
by $S_j\left( w,  \phi  \right)$, $j=1,2, \cdots s$, the set of
varieties, locus of the singularities of the integrand
($\phi$ denotes $\phi_1, \phi_2, \cdots, \phi_n$).
Landau equations amount to finding the necessary conditions for
the singularities to occur when the hypercontour is pinched between
the surfaces of singularities (pinch singularity) or meets a boundary
variety (end-point singularity).
For this to happen, the parameters $\alpha_i$ and $\beta_i$ should exist,
not all equal to zero, and such that at the point $(w, \phi)$
\begin{eqnarray}
\label{landaucond}
\alpha_j \cdot  B_j(w, \phi ) =0, \qquad  j= \,1, \,\cdots, \, d  \\
\beta_j \cdot  S_j(w, \phi ) =0, \qquad  j= \,1, \,\cdots, \, s \nonumber \\
 {\frac{\partial}{\partial \phi_j}} \Bigl( \sum_i \alpha_i \cdot  B_i(w,\phi)
+ \sum_i \beta_i \cdot  S_i(w,\phi)
 \Bigr) = 0, \qquad j= \,1, \, \cdots, \, n  \nonumber
\end{eqnarray}
Solving this set of equations will give the singularities $w$ that are
``candidates'' to be those of the integral (\ref{theintegral}).
Note that the easy resolution of these equations will depend on the number
of integration variables and the number of singularity varieties.

Consider, for instance, a one-dimensional integral in the form:
\begin{equation}
I_1\,\, =\,\,\,\int_{\,0}^{2\,\pi }\,{\frac{{d\theta }}{{D(\theta 
,\,w)}}}
\label{example1D}
\end{equation}
where $D(\theta ,\,w)$ represents the variety of singularities.
From (\ref{landaucond}), the Landau conditions read\footnote[5]{
Strictly speaking, one has to add $\beta (\theta -2\pi )=0$, but 
this equation does not change the discussion.
We are considering $2\pi$ periodic integrands.}: 
\begin{eqnarray}
\label{eqsLandau1} 
&&\alpha \cdot D(\theta ,\,w)\,=\,\,0,\,
 \qquad \quad \quad \beta \cdot \theta \,=\,\,0,\, 
 \nonumber \\
&&{\frac{{\partial }}{{\partial \theta }}}\,\Bigl(\alpha \cdot D(\theta
,\,w)\,+\,\beta \cdot \theta 
\Bigr)=\,\,0  
\end{eqnarray}
Knowing that $\alpha $ and $\beta $ should not be both equal to zero, 
it is easy to see that the set of equations (\ref{eqsLandau1}) gives 
two kinds of solutions, namely: 
\begin{eqnarray} 
\label{eqsLandau11} 
&& D(\theta ,\,w)\, =\,\,0,\qquad \quad {\theta }\,=\,\,0\quad
({\rm mod}\, \,  2\pi)   \\
\hbox{or:} && \quad
\label{eqsLandau12}
D(\theta ,\,w)\, =\,\,0,\qquad 
\quad {\frac{{\partial }}{{\partial \theta }}}\,D(\theta ,\,w)\,=\,\,0  
\end{eqnarray}
The first solution corresponds to {\em endpoint singularities} while 
the second one corresponds to {\em pinched singularities}. 
Note that mixed situations with  endpoint together with pinched singularities,
can occur for multidimensional integrals. 

\section{A multiple integral: the $y^2-$product}
\label{toy2}

Recall that a multidimensional integral of the product of
the $\,y_i$ quantities has been recently introduced by Nickel~\cite{nickel-05}
to suggest that the singularities, like $\, 1\,+3w\,+4w^2\,=\,0$,
appearing in~\cite{ze-bo-ha-ma-04} for the $\, \chi^{(3)}$ linear differential
equation are {\em pinch singularities}.

 In this section, for elegance, and
 less prohibitive calculations purposes, we consider, 
instead, a multidimensional integral of 
the product of $\, y_i^2$'s. The
results are similar to the product of the $\, y_i$'s, as far as the
singularities location is concerned (see Appendix A for the linear ODE 
bearing on the product of $y_i$ with $n=3$).

Our multidimensional integral thus reads:
\begin{eqnarray}
\label{defFN}
Y^{(n)}(w)\,\,  =\,\, \,  \,  \, 
 \int_0^{2\pi }\frac{d\phi _{1}}{2\pi }\, \int_{0}^{2\pi }
\frac{d\phi _{2}}{2\pi }\,\, ...\, \int_{0}^{2\pi }\frac{d\phi 
_{n-1}}{2\pi }
\left( \prod_{i=1}^{n} y_{i}^{2} \right)  
\end{eqnarray}
with
\begin{eqnarray}
\label{sumphi}
 \sum_{k=1}^{n} \phi_k \,=\, 0
\end{eqnarray}

To find the linear differential equations for these integrals, we make use
of the Fourier expansion of $y_{i}^{2}$ which reads (dropping the indices)
\begin{eqnarray}
\label{FourExp Y2}
&&\frac{y^{2}}{4w^{2}}\,  =\, \,\,  D(0)\,\, 
+2\, \sum_{k=1}^{\infty }\, D(k) \cdot \cos (k\phi )
\nonumber \\
&& \qquad =\,\, \sum_{k=-\infty }^{\infty }D(k)Z^{k} 
\,\,=\,\, \sum_{k=-\infty }^{\infty }\, D(k)\cdot Z^{-k}
\end{eqnarray}
with $\, Z=\, \exp (i \, \phi )$ and
\begin{equation}
D(k)\,=\,\, D(-k)=\,\,\,  w^{\left\vert k\right\vert } \cdot 
d(\left\vert k\right\vert )
\label{defD(k)}
\end{equation}
where $\, d(k)$ is a {\em non terminating} hypergeometric series given 
by
\begin{eqnarray}
\label{defdk}
&& d(k) \, =\,\,   \left( 1+k\right) \times \\
&& \qquad _{4}F_{3} \Bigl( \frac{3}{4}+\frac{k}{2},1+\frac{k}{2},
\frac{5}{4}+\frac{k}{2},\frac{3}{2}\, 
+\frac{k}{2};\frac{3}{2},1+k,\frac{3}{2}+k ;16\,w^{2} \Bigr)   
\nonumber
\end{eqnarray}
The Fourier expansion (\ref{FourExp Y2}), together with the definitions
(\ref{defD(k)}), (\ref{defdk}),
allow to write the expansion of the integral  $\, Y^{(n)}(w)$, fully 
integrated over the angles,
 as (see Appendix B for details)
\begin{eqnarray}
 \label{FN expansion}
Y^{(n)}(w) \, \,  =\, \, \, (4w^{2})^{n}\cdot 
\sum_{k=0}^{\infty } C_k \cdot w^{n\,k}
\cdot  d(k)^{n}  
\end{eqnarray}
where $\, C_0\,=\, 1$ and $\, C_k=\, 2$ for $\, k\,  \ge\,  1$.

\subsection{Linear ODE's and their solutions}
\label{ODEsol}
Series generation using (\ref{FN expansion}) can be carried out 
without any difficulty.
We have computed the linear differential equations satisfied by 
the $Y^{(n)}$'s up to $n=10$.
These Fuchsian linear ODE's have many remarkable properties.
They are of order \emph{two},
\emph{independently} of the dimension $n$ of the integral.
All these Fuchsian linear ODE's have the singularities 
$w=0$, $w=1/4$, $w=-1/4$ and $\infty $, with
respectively the critical exponents $(2n, n-(-1)^n)$, 
$\left(-(n-2)/2, -(n+1)/2\right)$,
$(-(n-2)/2, -(n+(-1)^n)/2)$ and $(-(n-1)/2, -(n+1)/2)$. The singularity
$w=0$ is an apparent one. These linear ODE's have 
other regular singularities with
critical exponents $-1$ or  $\, 0$.

Let us give, as an example, the linear ODE
for $\, n=3$ ($F(w)$ denotes $\, Y^{(3)}(w)$) 
\begin{eqnarray}
\sum_{m=0}^{2}\,a_{m}(w)\cdot {\frac{{d^{m}}}{{dw^{m}}}}\, 
F(w)\,\,=\,\,\,\,\,0
\end{eqnarray}
where
\begin{eqnarray}
&&a_2 (w)\,  = \,\, {w}^{2} \cdot \left( 1-w \right)  \, (1+3\,w)
 \left( 1-16\,w^2 \right)^{2}\nonumber \\
&& \qquad  \quad \quad  \quad \times  \left( 1-5\,w \right) 
 \left( 1+3\,w+4\,{w}^{2} \right) \cdot  P_2(w), \nonumber \\
&&a_1 (w) \,  = \, \, 
  \left( 1-16\,w^2 \right)\, w \cdot  P_1(w), \qquad \quad
a_0(w)\,  = \, \,  6\,P_0(w)  \nonumber
\end{eqnarray}
the polynomials being:
\begin{eqnarray}
&&P_2 \, = \,\,  
 1-2\,w-24\,{w}^{2}-5\,{w}^{3} +19\,{w}^{4}+98\,{w}^{5}
-574\,{w}^{6}-2835\,{w}^{7} 
 \nonumber \\
&& \qquad +2330\,{w}^{8}+392 \,{w}^{9},   \nonumber \\
&&P_1 \,  =\,\,   -9+16\,w+410\,{w}^{2}-8\,{w}^{3}-5164\,{w}^{4}
-7196\,{w}^{5}+12896\,{w}^{6} \nonumber \\
&&\qquad 
+65824\,{w}^{7}-192637\,{w}^{8}-1302860\,{w}^{9}+327886\,{w}^{10}
\nonumber  \\
&&\qquad  +6666128\,{w}^{11}  
+3469546\,{w}^{12}-6576560\,{w}^{13}\nonumber \\
&&\qquad  -2874016\,{w}^{14}+1967104 \,{w}^{15}+752640\,{w}^{16},  
\nonumber \\
&&P_0\,  =\,\,  
 4-6\,w-224\,{w}^{2}+29\,{w}^{3}+3808\,{w}^{4}+3272\,{w}^{5} -21680\,{w}^{6} 
 \nonumber \\
&& \qquad-71496\,{w}^{7}+87964\,{w}^{8}+1146670\,{w}^{9} -856344\,{w}^{10}
 \nonumber \\
&&\qquad -9773745\,{w}^{11}+2595240\,{w}^{12}
+30066076\,{w}^{13} +1626384\,{w}^{14}
 \nonumber \\
&& \qquad
-26569152\,{w}^{15} -8783616\,{w}^{16}-1204224\,{w}^{17} +2007040\,{w}^{18}.   
\nonumber
\end{eqnarray}
The roots of polynomial $\, P_2(w)$ are {\em apparent singularities}.
The linear ODE for $n=4$ is given in Appendix C.

This second-order linear differential equation can be solved exactly, 
and has the general solution
($c_1$ and $c_2$ are arbitrary constants):
\begin{eqnarray}
\label{solY3}
&&F(w) \, =\,  \,c_1 \cdot  F_1(w)  \, \,+ c_2 \cdot F_2(w),   \\
&&F_1(w) =\, \nonumber \\
&&\quad \quad 
{\frac { \left( 3-59\,{w}^{2}-82\,{w}^{3}+386\,{w}^{4}
+8\,{w}^{5}+64\,{w}^{6} \right)\cdot {w}^{4}  }{3\, \left( 1-4\,w 
\right)^{2} \left( 1-w \right)
  \left( 1+3\,w \right)  \left( 1+4\,w \right)  \left( 1-5\,w \right)
  \left( 1+3\,w+4\,{w}^{2} \right) }}, \nonumber \\
&&F_2(w) \,  =\, \, 
{\frac {\left( 1+2\,w-2\,{w}^{2} \right)\cdot {w}^{4}  }{ \left( 1-5\,w 
\right)  
\left( 1+3\,w+4\,{w}^{2} \right) \sqrt {1-16\,{w}^{2}} }}  
\nonumber 
\end{eqnarray}
Each component $\,F_i(w)$ is a solution of an
 order-one linear differential equation.
All the singularities are poles, except  $w = \pm 1/4$ which
are branch points for the second component $\,F_2(w)$.
For arbitrary constants of integration, the general solution 
$F(w)$ has the same singular behavior.
To obtain the particular solution corresponding to the integral 
(\ref{defFN}) for $\, n =\, 3$, i.e. $\,F(w)=\, Y^{(3)}(w)$, the 
constants of 
integration have to be fixed to 
$\, c_1=\, 12$, $\,c_2=\, -c_1$. 
Note that, when these constants of integration are fixed at these
values, the solution $\,F(w)=\, Y^{(3)}(w)$ is {\em no longer
singular} at $w=1/5$, and at the quadratic roots $w_0$ of
$1+3w+4w^2=0$, as can be easily checked:
\begin{eqnarray}
\label{1over5}
&&   Y^{(3)}(w) = \,\,\,  {37 \over 1100}, \quad 
\quad \quad \hbox{when} \qquad \quad
\quad w \rightarrow 1/5   \\
&&   Y^{(3)}(w) = \,\, \,  {4099 \over 90112}\,\, 
\pm i\, {3881\sqrt{7}\over 90112}, \, 
\quad \quad \quad \hbox{when}\quad \quad \quad w 
\rightarrow w_0  \nonumber 
\end{eqnarray}
However, these polynomials $\, 1-5\, w$ and  
$\, 1 +3\, w\, + 4\, w^2$ still remain in the expression of the solution.
Note that the corresponding singularities $\,1\, -5\, w\, =\, 0\, $
and $\,1\, +3\, w\, +4\, w^2\, =\, 0$ {\em are not} in ${\cal W}_c$,
while the others in (\ref{solY3}) are.

As a second example, consider, for instance, the case $\, n=\, 5$.
The singularities of the linear ODE,
besides $\, 1-16w^2=\, 0$, correspond to the roots of the two polynomials:
\begin{eqnarray}
\label{sets5}
&&Q^{(5)}_1 = \,    \left( 1-3\,w+{w}^{2}
 \right)  \left( 1+5\,w+5\,{w}^{2} \right),   \\
&&Q^{(5)}_2 =\,   \left( 1+5\,w+13\,{w}^{2}
 \right)  \left( 1-7\,w+5\,{w}^{2}-4\,{w}^{3} \right)\nonumber \\
&&\quad \quad  \quad    \left( 1-8\,w+20\,{w}^{2}-17\,{w}^{3} \right) 
\left( 1+8\,w+20\,{w}^{2}+15\,{w}^{3}+4\,{w}^{4} \right). \nonumber
\end{eqnarray}
One can check that the set of values of $\, w$ such that $1-16w^2\, =\, 0$, 
and the singularities of the set 
$\, Q^{(5)}_1=0$, are indeed singularities of the
 multiple integral $\, Y^{(5)}$ 
(see below and 
Appendix C),
while those of the set $\, Q^{(5)}_2=0$ are not.
Here also, the singularities cannot be "simplified" due to
the presence of the radical $\sqrt{1-16w^2}$.
One can also check that the singularities corresponding to
$\, Q^{(5)}_1=0$ are in ${\cal W}_c$, 
while those of the set $\, Q^{(5)}_2=0$ are not.
These remarkable facts hold for all the other $\, Y^{(n)}$'s.

With the differential equations up to $n=10$, we find that the 
singularities
are roots of the following polynomials 
given by Nickel \cite{nickel-05} 
\begin{eqnarray}
&& \cos \left( k \phi^{+} \right) - \cos \left( (n-k) \phi^{-} \right) 
\,\, = \,\, 0,  \\
\label{coscosnickel}
&& \cos \left( \phi^{+} \right)\,=\,\,1+{1 \over 2w}, \qquad 
\cos \left( \phi^{-} \right) \,=\,\, -1+{1 \over 2w}
\end{eqnarray}
which can be written as
\footnote[5]{Note that Nickel~\cite{nickel-05} excludes the cases $k=0$ 
and $k=n$.}:
\begin{eqnarray}
\label{landauYn}
T_k \Bigl( {\frac{1}{2w}}+1 \Bigr)\,  -T_{n-k} \Bigl({\frac{1}{2w}-1} 
\Bigr)\,  =\, \, 0,
\qquad k\, =\, 0,\, 1, \, \cdots,\,  n
\end{eqnarray}
where $T_k(w)$ are the Chebyshev polynomials of first kind.
Note that, in this model, the singularity $1+4w=0$ is not
given by this formula for $n$ odd.
From these singularities, {\em only those} corresponding to $k=\, 0$, 
and $k=\, n$, {\em are singularities} of the $\, Y^{(n)}$'s, i.e.
\begin{eqnarray}
\label{landauphys}
\Bigl( 1 \,  -T_{n} \Bigl({\frac{1}{2w}-1} \Bigr)\,\Bigr) \cdot 
\Bigl( 1 \,  -T_{n} \Bigl({\frac{1}{2w}+1} \Bigr)\,\Bigr)
  \, =\, \, \, 0.
\end{eqnarray}

We are, now, in position to write the generic structure of
the solutions $\, Y^{(n)}(w)$.
For $n$ odd, the solutions $\, Y^{(n)}(w)$ have the form:
\begin{eqnarray}
&&Y^{(n)}(w)\,\,\, = \\
&&\quad {\frac{{w}^{n+1}}{\left( 1-16\,{w}^{2} \right)^{n/2-1} \cdot 
Q^{(n)}_2}}
\cdot  \Bigl( 
{\frac {c^{(n)}_1\cdot
 R^{(n)}_1 }{ \sqrt {1-16\,{w}^{2}} \left( 1-4\,w \right) \cdot 
Q^{(n)}_1 }} 
 + c^{(n)}_2 \cdot  R^{(n)}_2 \Bigr) \nonumber
\end{eqnarray}

For $\, n$ even, they read:
\begin{eqnarray}
&&Y^{(n)}(w) \,\, =\,\, \,
{\frac{{w}^{n-1}}{
 \left(  \left( 1-16\,{w}^{2} \right)^{n/2-1} \right) \cdot  
Q^{(n)}_2(w)}}\, \,  \times 
\nonumber \\
&& \quad \quad \quad \quad \left( {\frac { c_1^{(n)}\cdot R^{(n)}(w) }{
 \left( 1+4\,w \right)^{3/2} \cdot  Q^{(n)}_1(w)  }}\, +{\frac {
 c_2^{(n)}\cdot  R^{(n)}(-w) }{ \left( 1-4\,w\right)^{3/2}\cdot  
Q^{(n)}_1(-w) }} \right) \nonumber
\end{eqnarray}
The coefficients $\,c_1^{(n)}$ and $\,c_2^{(n)}$ are integration 
constants.
The polynomials $R^{(n)}$, $R^{(n)}_1$ and $R^{(n)}_2$, in these
expressions, have rational coefficients and
are such that the constant coefficient in $\, w$ is equal to $\, 1$.
The polynomials $\,Q^{(n)}_1$ and $\,Q^{(n)}_2\, $ are constructed,
respectively,  from the roots of (\ref{landauphys}) and from the 
roots of the cumulative product (for  $k=1, \cdots, n-1$) of  
(\ref{landauYn}).
The set of roots $\,Q^{(n)}_1\, = \, 0$
(resp.  $\,Q^{(n)}_2\, = \, 0$) are
(resp. are not) in ${\cal W}_c$.

The general solutions above give the particular solution (\ref{defFN}) 
for $\, c_2^{(n)}=\, -c_1^{(n)}$. When the
 integration constants satisfy 
this condition
the solution is {\em no longer} singular at the roots of $\,Q^{(n)}_2$.

\subsection{Landau conditions for $Y^{(n)}$}
\label{Lancon}
Consider, now, the Landau conditions\footnote[8]{
See the resolution of Landau conditions for the more elaborate
family of integrals of Section (4).}
for these integrals $\, Y^{(n)}$
which amounts to solving
\begin{eqnarray}
\label{condY}
&& \alpha_j \cdot \Bigl( \left( {1 \over 2w}-\cos(\phi_j) \right)^2-1  \Bigr)
 \, = \,\,0,
\qquad j=1,2, \cdots, n,  \\
&& \beta_j \cdot \phi_j  \,=\,\,0, \qquad \qquad j=1,2, \cdots, n-1,
 \nonumber \\
&& 2 \, \alpha_j \cdot \left( {1 \over 2w}-\cos(\phi_j) \right) \cdot 
\sin(\phi_j) \nonumber \\
&& \quad  +2 \alpha_n \cdot \left( {1 \over 2w}-\cos(\phi_n) \right) \cdot 
 \sin(\phi_n)
+\beta_j\, =\,\, 0,
\quad  j=1,2, \cdots, n-1 \nonumber 
\end{eqnarray}
for $\alpha_j$ and $\beta_j$ not all equal to zero.
The Landau singularities are obtained for the following configurations.
In the first configuration, $\alpha_n=0$, one obtains the singularities
$w=\pm 1/4$.
The second configuration $\alpha_j \ne 0$, for all $j$, gives when at least
one $\beta_j$ is not zero, the singularity $w=1/4$. For
 $\beta_j=0$, for all $j$,
one obtains the equations (\ref{coscosnickel}) 
for, respectively, $k$ angles $\phi^{+}$ and $n-k$ angles $\phi^{-}$. 
The constraint (\ref{sumphi}) becomes:
\begin{eqnarray}
\label{phi+phi-}
k\, \phi^{+} \, + \, (n-k) \, \phi^{-} \, = \,\, 2\pi
\end{eqnarray}
which is equivalent to
\begin{eqnarray}
\label{cosY}
&& \cos\left( k\, \phi^{+} \right) \,=\,\cos\left( (n-k) \, \phi^{-} \right),
  \\
\label{sinY}
&& \sin\left( k\, \phi^{+} \right) \,=\,-\sin\left( (n-k) \, \phi^{-} \right). 
\end{eqnarray}
Using only the first condition (\ref{cosY}), one obtains 
(\ref{landauYn}) which are 
the singularities of the linear ODE, while using both conditions
(\ref{cosY}), (\ref{sinY}),
one obtains the polynomials corresponding to $k=0$ and $k=n$ in 
(\ref{landauYn}), which {\em only give the singularities of the 
multiple integral}.
The last configuration is embedded in the second configuration when
$\alpha_j = 0$ for $j=1, 2, \cdots, p$ and
$\alpha_j \ne  0$ for $j=p+1, \cdots, n$. These polynomials are
just given by (\ref{phi+phi-}) with $n-p$ instead of $n$, since,
in the constraint (\ref{sumphi}), the angles $\phi_j$,
$j=1, \cdots, p$ become equal to zero by the equations (\ref{condY}).
The Landau conditions for $Y^{(n)}$ give then the singularities
of $Y^{(p)}$, with $p=\, 1,2, \cdots, n-1$. These
singularities are {\em spurious} with respect to the linear ODE.

\subsection{Kramers-Wannier duality subtleties: $\, w$ versus  $\, s$}
\label{Kramers}

In the integrals of the previous section, the calculations are performed in
the variable $\, w$. In particular we have made no distinction
 between the high and low temperature
regime ($s$ small or large). When trying to bridge 
the families of ``Ising class'' integrals
we are presenting in this paper and the actual Ising model
 integrals $\, \chi^{(n)}$, 
we should point out the subtleties of using the self-dual variable $\, w$
which does not distinguish between the interior of the
unit circle $\vert s \vert <1$ and the exterior of the 
unit circle $\vert s \vert >1$.

Let us come back to  solution $Y^{(3)}$, and let us 
consider the singularity $\, 1 -5w =\, 0$. 
The solution $Y^{(3)}(w)$ is {\em not} singular at $w=\, 1/5$ 
where it takes a finite value (\ref{1over5}), although
the polynomial $1-5w$ still remains in the algebraic
expression of the function. One has a similar situation 
 for the singularities $\, 1\,+3w\,+4w^2\, =\, 0$.
The non "cancellation" of these polynomials, 
in the algebraic expression of the solution,
may mean that the term $\,1-16w^2$, occurring in the square
root (\ref{solY3}), is a perfect square when written in some
other variable.
The last is nothing else than 
the variable $s$.

The analytic continuation in the variable $s$ is performed by
writing the linear ODE corresponding to $\, Y^{(3)}$
in the variable $s$, and solving. One obtains the following solutions
\begin{eqnarray}
\label{dualY}
F(s)\, = \,\,\, \mu_1 \cdot Y^{(3)}(s) \, 
+ \, \, \mu_2 \cdot Y^{(3)}\Bigl({{1} \over {s}}\Bigr)  \qquad   \qquad 
\hbox{where}
\end{eqnarray}
\begin{eqnarray}
\label{Y3s}
&&Y^{(3)}(s) \,= \, \, {{N_Y} \over {D_Y}} \qquad  \quad  \quad 
\hbox{with:}   \nonumber \\
&&N_Y \, = \, \, \,  \left( 6\,{s}^{7}+{s}^{5}-11\,{s}^{4}
+8\,{s}^{3}-32\,{s}^{2}+16\,s-16 \right)\cdot {s}^{6},  \nonumber \\
&&D_Y \, = \, \, \, \left( 1+{s}^{2} \right) ^{2} \left( s+1
 \right) ^{2}\left( s-1 \right)^{4} \left( 2\,{s}^{2}+3\,s+2 \right) 
 \left( 2\,{s}^{2}-s+2 \right)\nonumber \\
&&\qquad  \qquad  \times \left( {s}^{2}+s+2 \right)  \left( s-2 \right)  
 \nonumber 
\end{eqnarray}

The polynomials $(\, 1\,-5 w \,)$ and $(\,1\, +3\, w\, +4\, w^2\,  \,)$, 
 written in the variable
$\,s$, factorize, respectively, as $\,(1-2s)\cdot (s-2)\, $ and 
$\, (1\,+ s \, + 2\, s^2) (2\,+ s \, + \, s^2)\, $.
The ''physical'' solution $\,Y^{(3)}(s)\,$ is {\em not} singular at the 
point $\,1\,-2s\, =\,0$ and at the quadratic points
 $\,1\,+ s \, + 2\, s^2\, =\,0$,  which are {\em inside the unit 
circle} ($|s| < 1$). 
Do note, however, that the ''physical'' solution 
$\,Y^{(3)}(s)\,$
{\em actually presents a singularity at} $\,s -2\, =\,0$, as well as 
at  $\,2\,+ s \, + \, s^2\, =\,0$, which are  
{\em outside the unit circle} ($|s| > 1$).
The same conclusions hold for $Y^{(n)}$, $n>3$.

Let us now present a second family of more elaborate
integrals for which we address the same questions:
do the Landau singularities identify with the
singularities of the linear ODE or with the
singularities of the integral ? Do
the subtleties encoutered using the self-dual 
variable $\, w$ instead of the $s$ variable
also occur in this family?

\section{The Diagonal model}
\label{Diag}

Let us first recall the fundamental, and deep, mathematical relation 
between Hadamard product of series,  multiple integration,
and diagonal reduction~\cite{lip-poo,Hadamard}.
Christol's conjecture amounts to saying
 that any algebraic power series in $\, n$ variables
is the ``diagonal''of a rational power series of $\, 2\, n$ 
variables~\cite{lip-poo,christol1,christol2,christol3}. 
These kind of quite remarkable
results~\cite{christol3,necer-97,fagnot-96,allouche-97} gave us
the idea to build our second ``toy model''
by considering the ``diagonal'' of the integrand (seen as a function
of  $(n-1)$ angles $\phi_j$) occurring in  the multiple integrals
corresponding to the $\, \chi^{(n)}$'s, and thus perform, instead
of $\, n-1$ integrals, {\em only one} integration.

At first sight, one can imagine that such a drastic reduction
completely trivializes the function one tries to calculate, and yields  
a new function with totally different singularities.
On the other hand, in view of the previous
non-trivial 
results~\cite{lip-poo,christol1,christol2,christol3,necer-97,fagnot-96,allouche-97},
one can, a contrario,  imagine that such a ``diagonal reduction'' 
procedure keeps ``some''
relevant analytical informations, especially on the singularities.
Replacing the quite involved multiple integral
by an integration on {\em  only one} angle,
and since the integrand is an algebraic function of trigonometric 
functions of
 this angle and of the variable
$\, w=s/(1+s^2)/2$, the result will be holonomic.
This integral on  only one variable will
be solution of a finite order linear differential equation
 in $\, w$ with polynomial coefficients.
Could it be possible that the singularities of this
(probably Fuchsian) linear ODE, keep some ``memory'' of the
singularities of the involved initial multiple integral?

The second set of integrals, we present in this paper, make use of the 
following simplifications.
From the integrand of $\chi^{(n)}$ in (\ref{chi3tild}),
quicking out $G^{(n)}$ and the product on $y_j$, 
we only keep the following quantity in the  integrand:
\begin{eqnarray}
\label{theratio}
{\frac{1}{n!}}\cdot   {\frac{1\, +\prod_{i=1}^{n}\, x_i}{1\,
 -\prod_{i=1}^{n}\, x_i}}\, \, \, 
=\,\,\,\,   -{\frac{1}{n!}}\,\,\,  +{\frac{2}{n!}}\cdot  {\frac{1}{1\, 
-\prod_{i=1}^{n}\, x_i}} 
\end{eqnarray}
This is reminiscent of the assumptions made in \cite{nickel-99} that 
the singularities (or a subset of them) arise from the symmetry points of 
the integrand and the vanishing of the denominator of the above quantity.

As previously explained, instead of performing 
the integration on $(n-1)$ angles $\, \phi_j$,
we integrate the above quantity on the principal diagonal, i.e. from a
$(n-1)$-dimensional integral we reduce ouselves to
 {\em  only one} integration angle 
$\,\phi\, $:
\begin{eqnarray}
\label{diagphi}
\phi_1=\, \phi_2= \, \cdots =\,  \phi_{n-1}=\, \phi, \quad
\qquad \phi_n\, =\, -(n-1)\, \phi
\end{eqnarray}

Call $\, \Phi_{D}^{(n)}$ the integral on $\phi$ of (\ref{theratio}):
\begin{eqnarray}
\label{chinaked}
\Phi_{D}^{(n)}\,=\,\,\,\,  
 -{{1} \over {n!}}\,\,  \,+ {{2} \over {n!}} \, \int_0^{2\pi} 
{\frac{d\phi}{2\pi}}
 \,   {\frac{1}{1\, -x^{n-1}(\phi)  \cdot  x ((n-1)\phi)}} 
\end{eqnarray}
where $\, x(\phi)$ is given by (\ref{thex}).  
Expression (\ref{chinaked}), when expanded in the variable $\, x$, 
becomes:
\begin{eqnarray}
\label{expchinaked}
\Phi_{D}^{(n)}\,\, =\,\,\, \,
-{{1} \over {n!}}\, \, \,  + {{2} \over {n!}} \,  \int_0^{2\pi} 
{\frac{d\phi}{2\pi}}
 \,   \sum_{p=0}^{\infty} \, x^{p(n-1)}(\phi) \cdot x^p ((n-1)\phi) 
\end{eqnarray}
The term $x^p$ has a Fourier expansion given by~\cite{ze-bo-ha-ma-05}:
\begin{eqnarray}
\label{xp}
x^p \, =\, \,\,  \, w^p \cdot  \Bigl( 
b(0,p) \,\,  + 2\, \sum_{k=1}^\infty
w^k \cdot   \cos{k\phi} \cdot  b(k,p) \Bigr)
\end{eqnarray}
where $b(k,p)$ is a {\em non terminating} hypergeometric series that reads
(with $m=k+p$): 
\begin{eqnarray}
&&b(k,p)\,=\,\,{{m-1 }\choose {k}} \times  \\
&&\quad \quad  {_4}F_3 \Bigl( {\frac{{(1+m)}}{{2}}}, 
{\frac{{(1+m)}}{{2}}}, {\frac{{(2+m)}}{{2}}},
 {\frac{{m}}{{2}}}; 1+k, 1+p, 1+m;\, 16w^2 \Bigr) \nonumber 
\end{eqnarray}
With this Fourier series, the integration rules are straigthforward.

The fully integrated expansion of $\Phi_{D}^{(n)}$ simply reads:
\begin{eqnarray}
&&\Phi_{D}^{(n)}\,\, =\,\, -{{1} \over {n!}}\, \, \,  \\
&&\quad \quad 
+{{2} \over {n!}} \,
 \sum_{p=0}^{\infty} \sum_{k=0}^{\infty}\, w^{n(k+p)}\cdot  C_k \cdot
 b(k,p) \cdot b\Bigl( k(n-1),p(n-1) \Bigr) \nonumber 
\end{eqnarray}
where $\, C_0=\, 1$ and $\, C_k=\, 2$ for $k \ge 1$.
These sums are clearly less consuming in the computational effort 
as compared
to the multidimensional integrals (\ref{chi3tild}).

\subsection{Fuchsian linear differential equations for $\Phi_D^{(n)}$ 
and their singularities}
\label{Fuchsian1}
Generating sufficiently large series in $w$ for each 
$\,\Phi_{D}^{(n)}$, 
we have actually been able to
find the corresponding linear differential equations up to $n=8$.
The linear ODE's for $\,\Phi_{D}^{(3)}$, 
and $\,\Phi_{D}^{(4)}$, are of order four
and are given in Appendix D.
For instance, the linear ODE satisfied by $\, \Phi_{D}^{(3)}$ reads:
\begin{eqnarray}
\label{yes}
\sum_{m=1}^{4}\,a_{m}(w)\cdot {\frac{{d^{m}}}{{dw^{m}}}} \, 
F(w)\,\,=\,\,\,\,\,0
\end{eqnarray}
where
\begin{eqnarray}
&&a_4 (w)  \,  = \, \left(1-w \right)  \left( 1+2w \right)  
\left(1-16w^2 \right)\,     \left( 1 \, +3 \, w \, +4 \, {w}^{2} 
\right)\,  {w}^{2} 
 \cdot P_4(w) \nonumber \\
&&a_3 (w) \,   = \, w \cdot P_3(w), \qquad
a_2 (w)  \,  =  \,  P_2(w), \qquad
a_1(w) =  \, P_1(w) \nonumber
\end{eqnarray}
the polynomials $\, P_i(w)$ being displayed in Appendix D.

One does remark, on this second example, that the singularities of the linear
ODE for $\chi^{(3)}$~\cite{ze-bo-ha-ma-04,ze-bo-ha-ma-05} 
actually pop out in the
linear ODE of this ``toy integral'',
 {\em in particular the quite non trivial} 
 $\,  1 \, +3 \, w \, +4 \, {w}^{2}\, = \, 0$ singularities.

The linear differential equation (\ref{yes}) is of
 Fuchsian type. It has just regular singularities:
one can actually see that the degree eleven
polynomial $\, P_4$ does correspond to {\em apparent} singularities.
The linear ODE's for $\,\Phi_{D}^{(3)}$ and for the next
$\,\Phi_{D}^{(n)}$'s, all have logarithmic singularities at the points
$\,w=\,0,\,\pm 1/4\, $ and $\,w=\,\infty$.
The other regular singular points have a $\,x^{-1/2}$ singularity.
The singularities $1-4w^2=0$  also carry  a $\,x^{1/2}$ behavior.
Note that the order $\, q$ of the linear differential equations of these
$\,\Phi_{D}^{(n)}$'s follows a simple (parity dependent) linear rule
as a function of $\,n$, namely 
$\,q \, =\, \,\left(2n \,- (-1)^n\, +9 \right)/4$.

Let us recall the ``nickellian'' singularities 
(\ref{Nick}) found by Nickel for 
the $n-$particle contributions $\, \chi^{(n)}$ to
 the magnetic susceptibility of Ising 
model~\cite{nickel-99}. These singularities, written as a 
function of Chebyshev polynomials of first kind
\begin{eqnarray}
\label{singNickel}
{ {1} \over {2\, w}} \, =\, \,  T_k \Bigl( \cos \left( {2\pi \over n} 
\right) \Bigr)\, 
+ \, T_m \Bigl( \cos \left( {2\pi \over n} \right) \Bigr), \qquad 0 \le 
k, m \le n
\end{eqnarray}
{\em can all be obtained from this ``toy'' model}. 
Recalling the definition of the variable $\, w=\, s/2/(1+s^2)$ which is 
self dual
(i.e. invariant by $s \rightarrow 1/s$), one may introduce the 
following partition on the singularities of the linear ODE
corresponding to $\Phi_D^{(n)}$.
We denote by $P_1^{(n)}$ the polynomials where the roots (i.e. 
singularities)
are given by (\ref{singNickel}).
 We denote by $P_2^{(n)}$ the polynomials where 
the roots are not given by (\ref{singNickel}) but  are 
in ${\cal W}_c$. Finally, we denote 
 by $P_3^{(n)}$ the polynomials where, at least, one root is 
not in ${\cal W}_c$.
Along this partition, the polynomials for $\,\Phi_{D}^{(n)}$,
$n=3, \cdots, 8$, are respectively (void means there is no
such type of polynomials):
\begin{eqnarray}
\label{partphi3}
 P_1^{(3)}\, &=& \, (1-w)(1+2w)(1-16w^2)\cdot w,  \\
 P_2^{(3)}\, &=& \, {\rm void}, \qquad  P_3^{(3)}\, = \, 1+3w +4w^2 
\nonumber
\end{eqnarray}
for  $\, \Phi_{D}^{(3)}$, 
\begin{eqnarray}
 P_1^{(4)}\,  &=& \,  (1-4w^2)\,(1-16w^2)\cdot w,  \\
 P_2^{(4)}\,  &=& \, {\rm void},  \qquad P_3^{(4)}\,  = \, {\rm void} 
\nonumber
\end{eqnarray}
for  $\, \Phi_{D}^{(4)}$, 
\begin{eqnarray}
 P_1^{(5)} \, &=&\,  (1+w)(1-16w^2)(1-3w+w^2)(1+2w-4w^2)\cdot  w, 
\nonumber \\
 P_2^{(5)} \, &=& \, (1-w)(1+2w), \\
 P_3^{(5)} \, &=& \, (1 -w -3w^2 +4w^3)(1 +8w +20w^2 +15w^3 
+4w^4)\nonumber
\end{eqnarray}
for  $\, \Phi_{D}^{(5)}$, 
\begin{eqnarray}
 P_1^{(6)} \, &=&\,  (1-w^2)\,(1-16w^2)\,(1-4w^2)\,(1-9w^2)\cdot w,  \\
 P_2^{(6)} \, &=&\,  {\rm void}, \qquad P_3^{(6)} 
\, =\,  1\,  -10\, w^2\,  +29\, w^4
\nonumber
\end{eqnarray}
for  $\, \Phi_{D}^{(6)}$,
\begin{eqnarray}
 P_1^{(7)} \, &=&\,  w \cdot (1-16w^2)(1-5w+6w^2-w^3)(1+2w-8w^2-8w^3) 
\nonumber \\
&& \,\, (1+2w-w^2-w^3), \nonumber \\
 P_2^{(7)} \, &=&\, (1+w)(1-3w+w^2)(1+2w-4w^2),  \\
 P_3^{(7)} \, &=&\, 
 (1+12w+54w^2+112w^3+105w^4+35w^5+4w^6) \nonumber \\
&& (1-3w-10w^2+35w^3+5w^4-62w^5+17w^6+32w^7-16w^8) \nonumber
\end{eqnarray}
for  $\, \Phi_{D}^{(7)}$,
\begin{eqnarray}
 P_1^{(8)} \, &=&\,   w \cdot 
(1-16w^2)(1-4w^2)(1-2w^2)(1-8w^2)(1+4w+2w^2) 
\nonumber \\
&& \,\, (1-4w+2w^2), \nonumber \\
 P_2^{(8)} \, &=&\,  (1-w^2)(1-9w^2),   \\
 P_3^{(8)} \, &=&\, 1-26w^2+242w^4-960w^6+1685w^8-1138w^{10} \nonumber
\end{eqnarray}
for  $\, \Phi_{D}^{(8)}$.

Discarding the singularities $\, 1-16w^2=\, 0$, one may remark, that for
$n$ odd, the singularities, roots of the polynomials
$P_2^{(n)}$, are also roots of the polynomials $P_1^{(n-2)}$.
The remark holds for $n$ even, with the additional feature that 
$1-4w^2=0$
are common singularities to $\Phi_D^{(n)}$, $n$ even.

\subsection{Singularities of the linear ODE's versus 
singularities of particular
solutions}
\label{SingversusSol}
Having not been able to find all the solutions of the linear 
differential equations
satisfied by (\ref{chinaked}), one can use the connection 
matrix method we introduced in~\cite{ze-bo-ha-ma-05c}
to find, for some values of $n$, the singular behavior 
of (\ref{chinaked}) at each regular singularity.

This connection matrix method~\cite{ze-bo-ha-ma-05c} was used to find 
the singular behavior 
of $\, \chi^{(3)}$ and $\, \chi^{(4)}$, the third and fourth 
``particle'' 
contribution to the
Ising magnetic susceptibility~\cite{ze-bo-ha-ma-05c}. Let us sketch the 
method.
The method consists in equating, at some matching
points, the two sets of series corresponding, respectively, to expansions
around a singular point and the nearest other singular point.
The matching points have to be taken in the radius of convergence of 
{\em both} series.
Connection matrices of two singularities that are
not ``neighbors'' have to be deduced using some path of ``neighboring'' 
connection matrices.

We denote  by $q$ the order of the linear differential equation. Connecting 
the local series-solutions 
at the regular singular points $w=\, w_1$,
and $\, w=\, w_2$, amounts to finding the $\, q \times q \, $ matrix  
$\, C(w_1, w_2)$
such that
\begin{eqnarray}
\label{Cz2n1}
{\cal S}^{(w_1)} \, \, =\,\,\,   C(w_1,w_2) \cdot {\cal S}^{(w_2)}
\end{eqnarray}
where $\, {\cal S}^{(w_1)}$ and $\,{\cal S}^{(w_2)}$ denote 
the vectors whose entries are the local series-solutions at 
respectively
the singular point $\, w_1$ and $\, w_2$.
These solutions are evaluated numerically at $q$ 
arbitrary points around a point 
belonging to both convergence disks of the series-solutions.
We have considered the linear differential equations corresponding to 
$\,\Phi_{D}^{(3)}$,
$\,\Phi_{D}^{(4)}$, $\,\Phi_{D}^{(5)}$ and $\,\Phi_{D}^{(6)}$.

For the toy integral $\,\Phi_{D}^{(3)}$ whose linear differential
equation has the
singularities $\,w=\,0$, 
$\,w=\,1/4$,
$\,w=\, -1/4$, $\,w=\,1$, $\,w=\,-1/2$, $\,w=\,\infty$ and $w=\, w_1,\, 
w_2$ where $\, w_1$ and
$w_2$ are the roots of $1+3w+4w^2=0$, one obtains the following
singular behavior around each singularity
\begin{eqnarray}
&&\Phi^{(3)}_D \left({\rm singular}, 1/4 \right)\,  =\,\,
   -{\frac{1}{4\pi}} \, \, \ln(x)\cdot S_3^{(1/4)}(x),  \nonumber \\
&& \Phi^{(3)}_{D} \left({\rm singular}, -1/4 \right)\,  = \, \,
 -{\frac{5}{8\, \pi}}  \cdot
 x \cdot  \ln(x)\cdot  S_3^{(-1/4)}(x), \nonumber \\
&&\Phi^{(3)}_{D} \left({\rm singular}, 1 \right) =\, 
 {\frac{i}{\sqrt{6}}}  \cdot  x^{-1/2}\cdot S_1^{(1)}(x),  \nonumber \\
&& \Phi^{(3)}_{D} \left({\rm singular}, -1/2 \right)\,  =\,\,   
\,{\frac{i}{\sqrt{3}}}  \cdot
\, x^{-1/2}\cdot S_1^{(-1/2)}(x), \nonumber \\
&&\Phi^{(3)}_{D} \left({\rm singular}, \infty \right) =\, \, 
\, {\frac{3\, i}{4 \, \pi}}  \cdot  x \cdot \ln(x) \cdot S_3^{(\infty)}(x), 
\nonumber \\
&& \Phi^{(3)}_{D} \left({\rm singular}, w_1 \right) =\, \, \, 
b_1 \cdot\, x^{-1/2} \cdot  S_1^{(w_1)}(x) \nonumber 
\end{eqnarray}
where $\, S_j^{(w_s)}$ are series-solutions analytical at $x=0$ ($x=1-w/w_s$ 
is the expansion local variable), and such that
$\, S_j^{(w_s)}(0)=1$. 
Remark the exact value of $\, b_1$, we obtain, for 
the singularities $1+3w+4w^2=0$, namely $\, b_1=\, 0$. 
Recalling the singularity partition (\ref{partphi3}), one sees that 
only the singularities given by the polynomials where all the roots
are in ${\cal W}_c$
are actually present in the integral.

We have also considered the linear ODE for
$\,\Phi_{D}^{(4)}$ which contains only singularities 
in  ${\cal W}_c$. All these singularities
occur in the integral.
The next example is the linear ODE for $\,\Phi_{D}^{(5)}$.
All the singularities, given as roots of the polynomial $P^{(5)}_1$,
occur in the integral, while the singularities roots of
$P^{(5)}_2$ and $P^{(5)}_3$ {\em do not}. Note that two complex singularities,
given by the polynomial $1 -w -3w^2 +4w^3=0$, and all the roots
of $1 +8w +20w^2 +15w^3+4w^4=0$ are not in ${\cal W}_c$.
The last example we have considered is $\,\Phi_{D}^{(6)}$. Its
 linear differential equation has four singularities
 out of ${\cal W}_c$, namely
 the four roots of  $1\, -10\, w^2\, +29\, w^4\, =\, 0$.
Here also, we found that  only the polynomials
where all the roots are in ${\cal W}_c$, 
{\em correspond to singularities of the integral}.

Note that the amplitudes at the singularities given by (\ref{singNickel})
can be obtained from~\footnote{This is equivalent
to the formula (14) forwarded by Nickel \cite{nickel-99} for
the $\chi^{(n)}$ and for the same type of singularities.}:
\begin{eqnarray}
{\cal A}\, = \, \, {\frac{2w_s}{n\sqrt{2(n-1)}}}\cdot 
\sum_{k,m} \, {\frac{\sin^2\left( {2\pi k \over n} \right)}
{\sqrt{1-\cos\left( {2\pi k \over n} \right)
 \cos\left( {2\pi m \over n} \right)  }}}
\end{eqnarray}
where the sum is restricted to the integers that satisfy
(\ref{singNickel}) for the singularity $w_s$.
This formula is in agreement with the results of the connection
matrix method, for $n=3, \cdots, 6$, confirming its efficiency.

\vskip 0.1cm
The natural question now, is to find out whether {\em the Landau 
singularities identify, or have an overlap, with
 the singularities of the integral 
or, rather,  with the singularities of the linear ODE}.

\subsection{Singularities from Landau conditions}
\label{SingfromLand}

Before solving the Landau conditions for our integrals
 $\, \Phi_D^{(n)}$, some additional comments
on  Landau  formulation should be underlined.
The quantity  $x_j$ given in (\ref{thex}), is written
as\footnote[1]{$i\,0^+$ is introduced for an easy control of our computations.}
\begin{eqnarray}
x_j \, =\,  {\frac{1}{2w}} -\cos (\phi _{j})\, 
+i\, \sqrt{1-\left( {\frac{1}{2w}} -\cos (\phi _{j})\right)^{2}+i\,0^+}  
\end{eqnarray}
Define
\begin{eqnarray}
\label{defangzeta}
&&\cos(\zeta)\, = \,\, {1 \over 2w}-\cos(\phi),  \quad \quad  \\
&&\sin(\zeta)\, = \, \, 
\sqrt{1-\left( {\frac{1}{2w}} -\cos (\phi _{j})\right)^{2}+i\,0^+}
\end{eqnarray}
and, similarly, $\cos(\zeta_n)$ and  $\sin(\zeta_n)$, where $\,\phi $ 
is changed to $\, (n-1)\, \phi$.

Using some identities on Chebyshev polynomials and noting that:
\begin{eqnarray}
 x^{n-1}(\phi)\,\, =\, \,\,  T_{n-1}\left(\cos(\zeta) \right)
\, +i\, \sin(\zeta) \cdot U_{n-2}\left(\cos(\zeta) \right)
\end{eqnarray}
the ratio (\ref{theratio}), with the
 constraint (\ref{diagphi}), is cast in the form
\begin{eqnarray}
{\frac{\cos(\zeta_n)\, +T_{n-1}\left(\cos(\zeta) \right)\, - i\, \sin(\zeta_n)
+i\, \sin(\zeta)\cdot  U_{n-2}\left(\cos(\zeta) \right) }
{\cos(\zeta_n)\, -T_{n-1}\left(\cos(\zeta) \right)\, - i\, \sin(\zeta_n)
-i\, \sin(\zeta)\cdot  U_{n-2}\left(\cos(\zeta) \right) }}
\end{eqnarray}
Some manipulations give for  
the integral $\, \Phi_D^{(n)}$:
\begin{eqnarray}
\label{phinratio}
 \Phi_D^{(n)} \,\,  = \, \,\,  \, {\frac{1}{n!}}\, \int_{0}^{2\pi } d\phi 
\cdot 
 {{ \sin(\zeta_n) \, -  \sin(\zeta) \,U_{n-2}(\cos(\zeta)) } \over 
{\cos(\zeta_n)\, -T_{n-1}(\cos(\zeta))}} 
\end{eqnarray}
where $\, U_n$ is the Chebyshev polynomial of second kind.
In this form the denominator of the integrand
is polynomial in $\, w$ and $\, \cos(\phi)$, resulting from a 
``rationalization''
of the integrand. This procedure introduces the ``Galois 
companions''
of the integral and we may expect the Landau conditions 
to generate, up to spurious singularities, all 
the singularities of the linear ODE rather 
than those of the integral.
The second comment is on the branch points appearing in the
numerator. The manifolds defining these singularities
should be added in the Landau conditions. 

In the sequel, we give the singularities, obtained from
the Landau conditions, in two cases.
Firstly we allow the production of the singularities of the
``Galois companions'', and consider the locus of singularities
of the branch points. In a second step, we restrict the analysis
on the integral without considering the branch points locus.

The singularities, when the branch points appearing in the
numerator are not considered in Landau conditions
(see Appendix E),
are the roots of: 
\begin{eqnarray}
\label{setL1}
 T_{n-1} \left( {\frac{1}{2w}}\,  -1 \right)\, \, +1\, 
-{\frac{1}{2w}}\,\, =\,\, 0  
\end{eqnarray}
and
\begin{eqnarray}
\label{setL2}
   T_{n-1} \left( {\frac{1}{2w}}\,  +1 \right)\,\, +(-1)^{n-1}\, 
-{\frac{1}{2w}}\, \, = \,\,0  
\end{eqnarray}
and the $\, w$-polynomials obtained by eliminating the variable $v$ 
from:
\begin{eqnarray}
\label{setL3}
&& T_{n-1} \left( v \right)\, \, 
+T_{n-1} \left( {\frac{1}{2w}}\, -v \right)\, - {\frac{1}{2w}}\, =\,\,  
0,   \\
&& U_{n-2} \left( v \right)\,\,  -U_{n-2}
 \left( {\frac{1}{2w}}-v \right)\,  = \, \,0 \nonumber 
\end{eqnarray}
When the branch points are included, we obtain the 
additional polynomials ($k$ and $m$ are integers),
given by:
\begin{eqnarray}
\label{setL1branch}
 T_{n-1} \left( {\frac{1}{2w}}\, -(-1)^k \right)\, + (-1)^m\, 
-{\frac{1}{2w}}\,\, =\,\, 0, 
\end{eqnarray}
and by the elimination of $\, v$ from:
\begin{eqnarray}
\label{setL2branch}
T_{n-1}(v) - \left( {\frac{1}{2w}}\, \pm 1 \right) \, =\, 0,  
\qquad  T_{n-1}\left( {\frac{1}{2w}}\, -v \right)\, \pm 1 \, = \, 0
\end{eqnarray}

Remarkably  we find that the roots of the polynomials (\ref{setL1}), 
(\ref{setL2}) and (\ref{setL3})
{\em identify with the singularities of the differential equations 
governing}
(\ref{chinaked}) up to $\, n=8$, i.e. with the singularities given in
the polynomials $\, P^{(n)}_1$, $\, P^{(n)}_2$ and $\, P^{(n)}_3$.
The check has been extented up to $\, n=\, 16$ with the 
singularities of
{\em the linear differential equations obtained, this time, modulo some
primes}.

Note that the polynomials given by the first set (\ref{setL1}) have all
their roots in ${\cal W}_c$, and similarly for the second set
(\ref{setL2}) for $n$ even.

The polynomials (\ref{setL1branch}), (\ref{setL2branch}) give 
additional singularities not given by (\ref{setL1}-\ref{setL3}).
For $n=3, 4, 5$, they read:
\begin{eqnarray}
 &&n \, =\,  3, \quad \quad   (1+3w)(1-5w)(1-w-4w^2) \nonumber \\
 &&n \, =\,  4, \quad \quad   
 \left( 1 \pm 3\,w-{w}^{2} \right) (1\, \pm 6\, w \, +8\, w^2 \, \pm 4 
\, w^3)  
\nonumber \\
 &&n \, =\,  5, \quad \quad   \left( 1+3\,w \right)
 \left( 4\,{w}^{3}+3\,{w}^{2}-w-1 \right)  \left( 5\,{w}^{2}+5\,w+1 
\right)
\nonumber \\
&&\quad \quad  \left( 1-16\,{w}^{2}-2\,{w}^{3}+56\,{w}^{4}\, 
-16\,{w}^{5}-63\,{w}^{6}+8\,{w}^{7}+16\,{w}^{8}
 \right)     \nonumber \\
&& \quad \quad \left( -1+8\,w-20\,{w}^{2}+17\,{w}^{3} \right).
 \nonumber 
\end{eqnarray}
These singularities do not appear in the corresponding linear differential 
equation, they are  {\em spurious}. 

Recall that these "spurious" singularities (with respect to the linear ODE) 
are obtained when the manifold of the branch points appearing in the 
numerator of (\ref{phinratio}) are considered in the Landau
conditions. Let us now, exclude these (spurious) conditions on the 
branch points and, furthermore, 
let us control the solutions 
introduced by the normalization procedure in order to extract
the "physical" Landau singularities\footnote[5]{By "physical", we mean
"concerns" the integral, i.e. a particular solution of the linear ODE.}
from the "Galois"
Landau singularities (see Appendix E). 
In this case, the singularities
are obtained as roots of the gcd of the polynomials:
\begin{eqnarray}
\label{newsetL1}
&& T_{n-1} \left( {\frac{1}{2w}}\,  -1 \right)\, \, +1\, 
-{\frac{1}{2w}}\,\, =\,\, 0,  \\
&&  1+  U_{n-2} \left({1 \over 2w}-1 \right) \,\, =\,\, 0   \nonumber
\end{eqnarray}
and
\begin{eqnarray}
\label{newsetL2}
&&   T_{n-1} \left( {\frac{1}{2w}}\,  +1 \right)\, +(-1)^{n-1}\, 
-{\frac{1}{2w}} \,\, =\,\, 0,  \\
&& \sqrt{1-\left( {1 \over 2w}-(-1)^{n-1} \right)^2 }+
\sqrt{1-\left( {1 \over 2w}+1 \right)^2 }\,U_{n-2}\left( {1 \over 2w}+1 
\right) \,\, =\,\, 0  \nonumber
\end{eqnarray}
and the $\, w$-polynomials obtained by eliminating the variable $v$ 
from
\begin{eqnarray}
\label{newsetL3}
&& T_{n-1} \left( v \right)\, \, 
+T_{n-1} \left( {\frac{1}{2w}}\, -v \right)\, - {\frac{1}{2w}}\, =\,\,  
0,  \\
&& U_{n-2} \left( v \right)\,\,  -U_{n-2} \left( {\frac{1}{2w}}-v 
\right)\,  = \, \,0,  \nonumber \\
&& \sqrt{1-\Bigl( {1 \over 2w}-T_{n-1}(v) \Bigr)^2}+
\sqrt{1-\Bigl( {1 \over 2w}-v \Bigr)^2}\cdot  U_{n-2} \left( {{1} \over  
{2w}} \, -v\right) \,\, =\,\, 0 \nonumber
\end{eqnarray}

We see that we obtain the same set of polynomials 
(\ref{setL1}), (\ref{setL2}) and (\ref{setL3}) 
dressed with constraints.
The singularities, obtained from these conditions, correspond
 to the roots of the following polynomials
\begin{eqnarray}
 &&n \, =\,  3, \quad \quad \quad    (1-4w)(1+2w)(1-w), \qquad \nonumber \\
 &&n \, =\,  4, \quad \quad \quad    (1-4w)(1+4w)(1+2w)(1-2w), \nonumber \\
 &&n \, =\,  5, \quad \quad \quad    (1-4w)(1+w)(1+2w-4w^2)(1-3w+w^2), 
\nonumber \\
 &&n \, =\,  6, \quad \quad \quad    (1-4w)(1+4w)(1-w^2)(1-9w^2)(1-4w^2), 
\nonumber \\
 &&n \, =\,  7, \quad \quad  \quad   (1-4w)(1-5w+6w^2-w^3)(1+2w-8w^2-8w^3) 
\nonumber \\
&& \qquad \quad \quad \quad  (1+2w-w^2-w^3), \nonumber \\
&& n \, =\,  8, \quad \quad  \quad   
(1-4w)(1+4w)(1-2w^2)(1-8w^2)(1-4w^2)(1-9w^2) 
\nonumber \\
&& \qquad \qquad \quad \quad  (1-4w+2w^2)(1+4w+2w^2) \nonumber
\end{eqnarray}
which are exactly (up to $\, 1+4w\, $ for $n$ odd) the roots
given by $P_1^{(n)}$, i.e. Nickel's for the Ising model 
(\ref{singNickel}).
All these conditions
(\ref{newsetL1}), (\ref{newsetL2}) and (\ref{newsetL3}) give the
``nickellian singularities'' obtained from the simple form given in 
(\ref{singNickel}).

Recall that we know exactly, by the results of the connection matrix 
method~\cite{ze-bo-ha-ma-05c} given in the previous section,
which singularities are actually present in the
 integrals for $n=\, 3,\, 4,\,  5\, $ and $n=\,6$,
and, especially, the {\em non occurrence} in the integrals of 
the singularities (not in ${\cal W}_c$) roots of $P^{(3)}_3$ 
for $\, n=3$, roots of   $P^{(5)}_2$,  $P^{(5)}_3$ for $n=5$
and roots of $P^{(6)}_3$ for $n=6$.
So, and at least for $n=3, 4, 5$ and $n=6$, the
singularities obtained as roots of the polynomials (\ref{newsetL1}), 
(\ref{newsetL2}) and (\ref{newsetL3}) are in
 complete agreement with the singularities
of the integrals. 

We thus see that, depending on how the Landau conditions are performed,
the Landau singularities may be a set larger than the singularities
of the linear ODE, or identify with the singularities of the ODE, or, even,
could be the singularities of the integral.

\subsection{The singularities of the linear ODE's: an involved set 
with a structure}
\label{involved}
Assuming that the polynomials (\ref{setL1}), (\ref{setL2}) and (\ref{setL3})
do indeed reproduce the singularities of the linear
ODE corresponding to the integrals $\Phi_D^{(n)}$ for $n > 16$, 
one may use these formula to see how the singularities accumulate in 
the
complex plane of the variable $s$, as $n$ goes larger.
Figure 1 shows the loci of the singularities outside 
the unit circle $\vert s \vert = 1$.

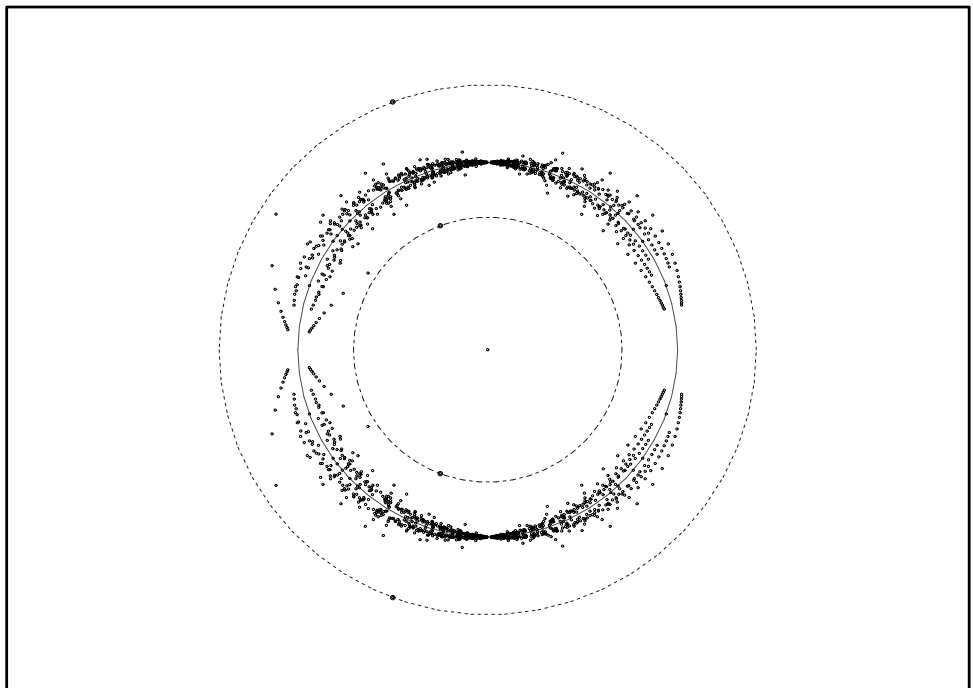
\image{Pn3singSq1}{-90}{0.98}{0.70}{The roots of $P^{(n)}_3$  
in the complex plane of the variable $s$, up to $n=47$.}

These singularities with $\vert s \vert \ne 1$ come only
from the polynomials (\ref{setL2}) for $n$ odd, and from the polynomials
(\ref{setL3}).
Note that the singularities from (\ref{setL2}) {\em are all
in the left half-plane of the variable} $s$.
The concentric circles are respectively (from the inner to the outer)
$\vert s \vert =1/\sqrt{2}$, $\vert s \vert =1$ and
$\vert s \vert =\sqrt{2}$.
The outer concentric circles 
 correspond to the modulus in $\, s$
of  the quadratic numbers
$\, 1\, +3 w\, +4 w^2=\, 0$ (that read in the variable $s$, $\, 
(1+s+2s^2) (2+s+s^2) =\, 0$).
We note the remarkable fact that all the singularities are lying
in the annulus defined by these inner and outer circles and are far from
the singularities $s=\pm 1$.
These last two points seem to be accumulation points to which
the singularities are converging as $n$ goes higher.

The proliferation of the singularities out of the unit circle $\vert s 
\vert = 1$ is
obviously polynomial. In fact, the degree of the 
polynomials  $\, P_3^{(n)}$ follows the simple rule
\begin{eqnarray}
\label{degreeP}
 {\rm deg}\left( P_3^{(n)} \right)\, =\, \, {1 \over 8} \left( 2n^2-15 
\right)
 - {(-1)^n \over 8} \left( 4n+1 \right), \quad \quad n=3,4, \cdots 
\nonumber 
\end{eqnarray}

When ``all'' the solutions of the ODE for each $n$ are considered,
their singularities form the unit circle $\vert s \vert=1$, and the
structure around, shown in Fig. 1. 

\subsection{More Kramers-Wannier duality subtleties: 
analytical continuation in $s$}
\label{MoreKra}
Let us consider, now, the linear differential equations
corresponding to the integrals $\Phi^{(n)}_D$, which, 
written in the variable $\, s$, are  obviously covariant
by Kramers-Wannier duality $s \leftrightarrow 1/s$.
Note that, this {\em does not mean} that the solutions of these self-dual
differential equations are invariant by 
Kramers-Wannier duality $s \leftrightarrow 1/s$.
Consider first, the simplest integral of this family,
namely  $\Phi^{(2)}_D$. As a function of the variable $\, w$ 
it is a solution of a third order differential operator which is 
the direct sum of two linear differential operators:
\begin{eqnarray}
\label{directw}
 Dw \, \oplus\Bigl(  Dw^2 \, 
+{{1\, -48\, w^2 } \over {(1-16\, w^2) \, w }} \cdot Dw \, 
\, -{{16 } \over { 1-16\, w^2 }}\Bigr)
\end{eqnarray}
with general solution
\begin{eqnarray}
c_0\,\, + c_1 \cdot K(4w)\,\,  + c_2 \cdot CK(4w)
\end{eqnarray}
where ($CK$) and $K$ are respectively the (complementary) complete
elliptic integral of first kind:
\begin{eqnarray}
\label{complete}
&& CK(4w) \, \,\,  = \, \,\, K\Bigl(\sqrt{1-16\, w^2}\Bigr) \\
&& K(4w) \, \,\,  = \, \,\,
 _{2}F_1\Bigl([1/2,1/2],[1],\, 16 \, w^2 \Bigr)
\end{eqnarray}
The linear differential operator (\ref{directw}),
 written in the variable $\, s$, gives
 a third order linear differential operator $\,  L_3(s)$ which reads:
\begin{eqnarray}
\label{directs}
&& L_3(s)  \, = \, \,   Ds \, \oplus L_2(s) 
\qquad \quad \hbox{where,}   \nonumber \\
&&  L_2(s)  \, = \,\,\, Ds^2 \, 
+{{s^4 +4 \, s^2 \, -1 } \over {(s^4\,  -1) \, s}} \cdot Ds \, 
\, -{{4 } \over { (1+s^2)^2 }} 
\end{eqnarray}
The third order linear differential operator
 $\,  L_3(s)$ has the following general solution:
\begin{eqnarray}
c_0\,\, + c_1\cdot  (1+s^2)\cdot  K(s^2)\, \,
 + c_2\cdot  \left( 1+ s^2 \right) \cdot 
CK \left( s^2 \right)
\end{eqnarray}
The corresponding linear differential
operators associated with the $\Phi^{(n)}_D$ have
 the constant fonction as solution. Note that,
 for the integrals $\Phi^{(n)}_D$,
this constant\footnote[8]{We discard from
now on, the factor $1/n!$ in the definition of the
integrals $\Phi^{(n)}_D$.} can actually be calculated and is equal to $1/2$.

The particular solution  $\Phi_D^{(2)}(s)$ reads:
\begin{eqnarray}
\Phi^{(2)}_D(s) \, =\,\, \, {1 \over 2}\, 
+ \, \,\,  {1 \over 2} (1+s^2) \cdot  K(s^2)
\end{eqnarray}
The general solution of $L_2(s)$ can be written as:
\begin{eqnarray}
\mu_1 \cdot \Phi^{(2)}_D(s)\, 
+\,\,  \mu_2 \cdot \Phi^{(2)}_D \left( {1 \over s} \right)
\end{eqnarray}
The two solutions of $L_2(s)$ are {\em actually the
 same function} for, respectively, the argument  $\, s$ and  $\, 1/s$.
Note that, this can be easily seen on 
the {\em formal solutions} of $L_2(s)$ which
take the same form
\begin{eqnarray}
&& \, \Phi^{(2)}_D(u)-1/2,  \nonumber \\
&& \left( \Phi^{(2)}_D(u)-1/2 \right)\cdot  \ln(u) \,\, + {1 \over 16}
\left( u^4+u^6+{21 \over 32}u^8 + \, \cdots \right)
\end{eqnarray}
where $u=\, s$ at the singular point $s=\, 0$ and 
$u=\, 1/s$ at the point at the
infinity.

The two quantities $\, \Phi^{(2)}_D(s)$ and $\, \Phi^{(2)}_D(1/s)$
are two different solutions of the self-dual differential
equations, in complete analogy with $\, Y^{(3)}(s)$ and
$\, Y^{(3)}(1/s)$ of Section (3.3), especially
 their singularities which clearly break 
$s \leftrightarrow 1/s$ symmetry.

We have here the exact equivalent of the 
situation we had~\cite{ze-bo-ha-ma-05b}  with
the order-two linear differential equation corresponding to
$\, \tilde{\chi}^{(2)}$. The latter reads:
\begin{eqnarray}
\label{chi216w2}
 \tilde{\chi}^{(2)}\,  = \,\,\,
 4 \,w^{4}\cdot {_{2}}F_{1}\Bigl([5/2, 3/2], [3],  \, 16\, w^2 \Bigr)
\end{eqnarray}
while the second independent solution~\cite{ze-bo-ha-ma-05b} reads
in terms of the MeijerG function~\cite{Meijer}:
\begin{eqnarray}
\tilde{{\cal S}}_3 (w) \, = \,\,\, {\frac{\pi}{2}}\,
{\rm MeijerG} \left( [[],[1/2,3/2]], [[2,0],[]],16w^2 \right)
\end{eqnarray}
Rewriting the order-two linear differential equation in the variable $s$, 
one obtains the general solutions:
\begin{eqnarray}
\label{solving}
&&F(s)\,  = \,\,\,
 c_1 \cdot {1 \over s^4}\cdot  \left(1+{1 \over s^2} \right)
 \cdot {_{2}}F_{1}\Bigl([5/2, 3/2],  [2],  {1 \over s^4} \Bigr) \nonumber \\
&& \qquad  \qquad  \qquad  +\,  c_2 \cdot  s^4 \cdot \left(1+ s^2 \right)
 \cdot {_{2}}F_{1}\Bigl([5/2 ,3/2], [2],   s^4 \Bigr) \nonumber \\
 &&\qquad =\,\,\,\, 
c_1 \cdot   \tilde{\chi}^{(2)}(1/s)\, + c_2 \cdot  \tilde{\chi}^{(2)}(s)
\end{eqnarray}

The compatibility of (\ref{chi216w2}) and (\ref{solving})
corresponds to remarkable identities like:
\begin{eqnarray}
\label{crucial1}
 &&    _{2}F_1\Bigl([5/2,3/2],[3],\, {{ 4 \, s^2} 
\over {(1+s^2)^2}}\Bigr)\, \,\, = \, \,\, \\
&&\qquad \qquad  \,\, = \, \,\, 
 (1+u^2)^5\,  \cdot \, _{2}F_1([5/2,3/2],[2],\, u^4) \nonumber 
\end{eqnarray}
or
\begin{eqnarray}
\label{crucial2}
 && _{2}F_1\Bigl([1/2,1/2],[1],\, {{ 4 \, s^2} 
\over {(1+s^2)^2}}\Bigr)\, \,\, = \, \,\, \\
&&\qquad \qquad  \,\, = \, \,\, 
 (1+u^2)\, \cdot \, _{2}F_1([1/2,1/2],[1],\, u^4) \nonumber 
\end{eqnarray}
where $u=s$ for small values of $s$ and $u=1/s$ for large values of $s$.

The previous calculations deal with simple linear ODE's and
simple enough functions. However,
for solutions of Fuchsian linear differential equations of higher order
it becomes a wishful thinking 
to recognize, in the whole $\, s$-complex plane, that two functions 
are the ``same function'' for, respectively, the arguments
$\, s$ and  $\, 1/s$.

Consider, for instance, the  integral $\, \Phi^{(3)}_D(s)$.
The order four linear differential operator for $\Phi^{(3)}_D(s)$ has the
following direct sum decomposition:
\begin{eqnarray}
\label{directsum}
Ds \oplus {\cal L}_3(s)\,   = \,\,\,  
 Ds \oplus \Bigl({\cal L}_2(s) \cdot  {\cal L}_1(s) \Bigr)
\end{eqnarray}
where the order-one, and order-two, linear differential operators
are given in Appendix F.

The integral $\Phi^{(3)}_D(s)$ is the sum of the constant $\, 1/2$ and of
a particular solution of ${\cal L}_3$:
\begin{eqnarray}
\label{int}
S({\cal L}_3) \,  =\, \, \, \,  c_1 \cdot  S({\cal L}_1)\,\, 
  +S({\cal L}_1) \cdot \int{ds\, {\frac{S({\cal L}_2)}{S({\cal L}_1)}}}
\end{eqnarray}
where $S({\cal L}_1)$ is the solution of the 
self-dual order-one  linear differential
operator ${\cal L}_1$:
\begin{eqnarray}
S({\cal L}_1)\, =\,\, 
{\frac { \left( 1+{s}^{2} \right)  \left( 2+s+2\,{s}^{2} \right) }{
\sqrt {1+s+{s}^{2}}\, \sqrt {2-s+2\,{s}^{2}}\, \sqrt {2+s+{s}^{2}}\, \sqrt {1+s
+2\,{s}^{2}}}}
\end{eqnarray}
The general solution of the order-two linear differential operator ${\cal L}_2$
 is a linear combination of  $\, S_1({\cal L}_2)$
and  $\, S_2({\cal L}_2)$ given by:
\begin{eqnarray}
\label{upto}
&& S_1({\cal L}_2)(s)\,  = \,\, \, 
{1 \over D_0} \cdot \Bigl( D_K \cdot  K(s^2) + D_E \cdot  E(s^2) \Bigr),  \\
&& S_2({\cal L}_2)(s)\,  =\, \, \,\,   {1 \over s^2}\cdot  S_1({\cal L}_2)(1/s) \nonumber
\end{eqnarray}
where $\, D_0$, $\, D_K$ and $\, D_E$ are polynomials in $s$ given
in Appendix F,
and $E(s^2)$ is the complete elliptic integral of second kind.
The order-two linear differential operator ${\cal L}_2$ is not self-dual, it is 
self-dual ``up to $\, s^2$'' (see (\ref{upto})).

To see that $\Phi^{(3)}_D(s)$  and $\Phi^{(3)}_D(1/s)$ are two different
solutions of the linear differential operator (\ref{directsum}), the
integration in (\ref{int}) has to be worked out.

As far as the singular behavior of
solutions of finite order linear 
differential operators with polynomial coefficients is concerned, 
 the connection matrix method we introduced in~\cite{ze-bo-ha-ma-05c}
{\em actually provides an answer to these difficult problems of
analytical continuation in the  whole $\, s$-complex plane}.
For the singular behavior of $\, \Phi^{(3)}_D(s)$ at the points
$\, (2+s+s^2)(1+s+2s^2)=0$,
we have, typically, to compose three
$3 \times 3$ connection matrices linking the point $s=0$
to $s_1=-1/4+i\,\sqrt{7}/4$, root of $1+s+2s^2=0$ which is inside
the unit circle,
then to $s=i$, then to $s_2=-1/2+i\,\sqrt{7}/2$, root of $s^2+s+2=0$
which is outside the unit circle.
The corresponding singular behavior are:
\begin{eqnarray}
\label{singul}
&& \Phi^{(3)}_{D} \left({\rm singular}, s_1 \right) =\,\, 
b_1 \cdot\, x^{-1/2} \cdot  S_1^{(s_1)}(x),  \nonumber \\
&& \Phi^{(3)}_{D} \left({\rm singular}, s_2 \right) =\, \,
b_2 \cdot\, x^{-1/2} \cdot  S_2^{(s_2)}(x) 
\end{eqnarray}
where $\, S_j^{(s_j)}$ are series-solutions, analytical at $x=0$ 
($x$ is the expansion local variable $x=\, 1-s/s_i$),
and such that
$\, S_j^{(s_j)}(0)=1$. The amplitudes at the singularities are
$b_1\, =\, 0$ and $\, b_2=\, (1+i)\cdot 7^{-1/4}/2$.
This means that the integral $\Phi^{(3)}_D$ is {\em not} singular 
at  the two quadratic roots $\,1+s+2s^2\,=\, 0$ 
which are inside the unit circle 
($\vert s \vert <1$). In contrast $\Phi^{(3)}_D$ actually exhibits a singular behavior
at the two quadratic roots $\,2+s+s^2\,=\, 0$ which are outside 
the unit circle ($\vert s \vert\,>\,1$).

We have also considered the singular behavior of $\Phi^{(5)}_D$ in the
complex plane of the variable $s$. The singularities where at least
one is out of the unit circle ($\vert s \vert\,\ne \,1$) 
are given by the roots of the polynomial
$P^{(5)}_3$ which decomposes in two polynomials reading, respectively,
$(4-2s+9s^2-2s^3+9s^4-2s^5+4s^6)$ and
$(4+6s+7s^2+5s^3+2s^4)(2+5s+7s^2+6s^3+4s^4)$.
The integral $\Phi^{(5)}_D$ is not singular at the roots of the
first polynomial and at the roots of $(2+5s+7s^2+6s^3+4s^4)=0$,
while its analytic continuation is {\em actually singular} at the roots of
$(4+6s+7s^2+5s^3+2s^4)=0$ which are exterior to the unit circle ($|s| >1$).
For  $\Phi^{(6)}_D$, the ODE carries 
as singularities out of the unit circle, the roots of
$1\,-10w^2\,+29w^4=\,0$ which reads in the variable $s$,
$16 +24 s^2\,+45s^4\,+24 s^6\,+16 s^8=\,0$. The integral $\Phi^{(6)}_D$
is {\em not singular} at these points.

As a result, one may imagine that the singularities
of the $\, \Phi^{(n)}_D(s)$, defined in the whole complex
plane of the variable $s$ by analytical continuation
of high temperature $s-$series (valid for $\vert s \vert <1$),
actually correspond to the ``nickellian singularities'' (\ref{Nick}), 
{\em together with some points in the  "cloud" of singularities} of Fig. 1
outside the unit circle ($\vert s \vert >1$).

In view of these results on the integrals
$\, Y^{(n)}$ and $\, \Phi^{(n)}_D$, it is necessary
to revisit, {\em in the $\, s$ variable},  the connection matrix computation 
for the third contribution to the susceptibility
of Ising model, $\tilde{\chi}^{(3)}$ given
in~\cite{ze-bo-ha-ma-05c}.

We find that, similarly, to our toy integral $\, \Phi^{(3)}_D$,
$\tilde{\chi}^{(3)}$ is {\em not singular} at the two roots $\, 1+s+2s^2\, =\, 0$, 
for which $\vert s \vert < 1$, while its analytical continuation is 
{\em actually singular} at the two roots $2+s+s^2 =\, 0$, for which $\vert s \vert >1$.
The corresponding singular behavior reads:
\begin{eqnarray}
&& \tilde{\chi}^{(3)} \left({\rm singular}, s_1 \right)\, =\,\, 
b_1 \cdot\, S_1^{(s_1)}(x)\cdot x \cdot \ln(x), \nonumber \\
&& \tilde{\chi}^{(3)} \left({\rm singular}, s_2 \right)\, =\, \,
b_2 \cdot\,  S_2^{(s_2)}(x)\cdot x \cdot \ln(x) \nonumber 
\end{eqnarray}
where $\, S_j^{(s_j)}$ are series-solutions, analytical at $x=0$ 
($x$ is the local variable  $x=1-s/s_i$)
and such that $\, S_j^{(s_j)}(0)=1$. 
 The amplitudes at the singularities are
$\, b_1= \, 0$ and $\, b_2\, =\,(181\sqrt{7}\, -7\, i)/256\pi$.

The three-particle contribution~\cite{ze-bo-ha-ma-05c} to the susceptibility
of Ising model, $\chi^{(3)}(s)$, is fundamentally 
a {\em high-temperature quantity}. The three-particle
 contribution, $\, \chi^{(3)}(s)$, together with 
$\chi^{(3)}(1/s)$, are {\em two different solutions} of 
the self-dual linear differential operator for $\chi^{(3)}(s)$.

Similarly to the $\, \Phi^{(n)}_D(s)$, one may imagine that the singularities
of the $\, \chi^{(n)}(s)$, for $n$ odd, defined in the whole complex
plane of the variable $s$ by analytical continuation
of high temperature $s-$series (valid for $\vert s \vert <1$),
actually correspond to the ``nickellian singularities'' (\ref{Nick}), 
{\em together with  a similar "cloud" of singularities} as in Fig. 1
outside the unit circle ($\vert s \vert >1$).

\subsection{Analytic continuation of Landau conditions in the variable $s$}
\label{Landauins}
These Kramers-Wannier duality subtleties, and the breaking of
the $s \leftrightarrow 1/s$
symmetry, can also be seen on the Landau conditions.
Consider for instance, the polynomial $1+3w+4w^2=0$ which appears
in the linear ODE for $\, \Phi^{(3)}_D$ from the condition (\ref{setL2}) and which does not occur
in the integral due the constraint on this condition given
in (\ref{newsetL2}).

When one switches to the variable $s$, the polynomial
$(1+3w+4w^2)$ becomes $(1+s+2s^2)(2+s+s^2)$. The constraint in
(\ref{newsetL2}) gives:
\begin{eqnarray}
\label{condLand}
&&\sqrt {-{\frac { \left( 1+{s}^{2} \right)  \left( s-1 \right) ^{2}}{{s}^{2}}}}\cdot s\,  \\
&& \qquad \qquad  \qquad \, +\, 2\,\sqrt {-{\frac { \left( 1+{s}^{2} \right) 
 \, (s+1)^2}{s^2}}} \cdot  (1+s+s^2)\,\, =\,\,\, 0\,   \nonumber 
\end{eqnarray}
The series expansions of the left-hand side of the algebraic
condition (\ref{condLand}) identify with
the  series expansions of:
\begin{eqnarray}
{{ \sqrt {1+{s}^{2}}} \over {s}} 
\cdot \left( 1+2\,s \right)  \cdot \left( 2+s+{s}^{2} \right)\,=\,\, 0 
\end{eqnarray}
and
\begin{eqnarray}
{{ \sqrt {1+{s}^{2}}} \over {s}} \cdot \left( s+2 \right) \left( 2\,{s}^{2}+s+1 \right)\, =\,\,0
\end{eqnarray}
for respectively, small values of $s$ and large values of $s$.
The gcd in (\ref{newsetL2}) written in $s$, and for small values of
$s$,  then gives the two quadratic roots
$2+s+s^2=\, 0$ as singularities of the integral $\Phi_D^{(3)}$.
These singularities are in the exterior of the unit circle ($|s| > 1$).
Similarly, had the integral $\, \Phi_D^{(3)}$ been defined for 
the large values of $s$, the gcd gives, as singularities,
the roots of the polynomial
$ 2\,{s}^{2}+s+1 =0$, which are in the interior $|s| < 1$ of the unit circle.

\section{Comments on the resolution of Landau conditions}
\label{comments}
In Section (2), we have recalled some basic general ideas 
and definitions on the Landau conditions.
We have used these conditions on 
our two families of integrals, for which, we knew, from the outset,
the set of singularities occurring in the linear ODE and in the integral.
In practice the calculations
can be slightly more subtle, and require cautious\footnote[9]{See for 
instance the examples in Itzykson and Zuber's book~\cite{itz-zub-80}, or
in Eden {\it et. al}~\cite{Smatrix}.}, sometimes tricky and 
involved, analysis
 on one (resp. several) complex variable(s).
For instance, the condition (\ref{sinY}) in Section (\ref{Lancon}) which
rules out all the singularities not occurring in the integral,
has to be worked out carefully to be able to control the
$\pm$ signs floating around.

The idea of the Landau conditions is to get some candidates for the
singularities of some (multiple) integral 
from simple enough (algebraic) calculations on the integrand.
In practice when the 
integrands are not rational expressions but algebraic ones, 
the algebraic character of the integrand introduces some ambiguity
for each of the branchs cuts.

For the calculations where the algebraic
expressions are quite involved, 
the control of the Riemann sheet we stay on,
may be quite tedious, sometimes hopeless.
The algebraic integrand being solution of some polynomial,
possibly of large degree, a possible 
approach amounts to considering all these roots together, 
namely the integrand of the 
actual ``physical'' (multiple) integral we are interested in, together 
with all his Galois ``companions''.
Recall that, as a consequence of the algebraic character of these
integrands, the (multiple) integral 
we are analyzing is holonomic. It is a solution of a finite order 
linear differential equation with polynomial coefficients, and it is therefore,
canonically associated with its ``Galois companions'' with 
respect\footnote[8]{The Galois group of permutation
of the roots of the polynomial is replaced here by the differential 
Galois group of the 
 linear differential equation.} to the linear ODE, namely the other 
solutions of the linear ODE. 

The calculations of the Landau conditions can now be 
performed for all these roots, leading to a larger set of Landau 
singularities that are probably good candidates for, not only
the ``physical'' (multiple) integral we are interested in,
but also for other mathematical expressions
corresponding to (multiple) integral of his  Galois ``companions'' 
integrands.

The conditions, we have used to settle the set of singularities
occurring in the integrals from the full set of singularities of
the linear ODE, appear in (\ref{cosY}), (\ref{sinY}) for the first family,
and in (\ref{eqcos}), (\ref{eqsin}) for the second family.
Note that (\ref{eqcos}) is the denominator appearing in 
(\ref{phinratio}) obtained with a ``rationalization procedure'',
introducing the "Galois companions" of the integrand, and
thus leading to the singularities of the linear ODE.
We are aware that, for many examples (not of the Ising class we
are interested in),
the resolution of Landau conditions may be more tricky and/or
more subtle.

Let us close this section by a comment on the
$w$ versus $s$ analytic continuation.
We have used the self-dual variable $w =\, s/2/(1+s^2)$ in our
previous works~\cite{ze-bo-ha-ma-04, ze-bo-ha-ma-05, ze-bo-ha-ma-05b, ze-bo-ha-ma-05c}
on the susceptibility of Ising model since
it appears naturally in the integrand (\ref{chi3tild})
and, especially, it is much simpler as far as computational efforts are concerned.
The differential equation for $\, \chi^{(3)}$ has been
obtained~\cite{ze-bo-ha-ma-04} with a series up to $\, w^{359}$.
In the variable $s$, a series
up to $\, s^{699}$ would have been needed.
When the distinction between the regimes of large values
of $s$, and small values of $s$, is required, we have seen some of the subtleties that pop out.
For the $\, Y^{(n)}$ family, this subtlety appeared through
a perfect square mechanism in the complex variable $s$.
For the $\Phi^{(n)}$ family, an equivalent mechanism of
factorization appeared and the result depends on the regime
considered (small $s$, large  $s$).
Note that this may not be the only mechanism that can be at work.
One may imagine, for instance, a function, like the difference
between the right-hand-side and the left-hand-side of (\ref{crucial1}),
that evaluates differently according to the region
 in the complex plane of the variable $\, s$ we are considering.

\section{Towards Ising model integrals}
The two families of integrals presented in the previous sections are
very rough approximations to the integrals (\ref{chi3tild}). 
The first family considered the product on $y_i$,
integrated on the whole domain of integration of the
$\phi_i$. Here, we found a set of singularities occurring in 
the $\chi^{(n)}$ and the quadratic polynomial $1\, +3w\, +4w^2=\, 0$.
The second family was constructed, besides discarding the factor
$G^{(n)}$ and the product on $y_i$, by restricting the domain of
integration on the principal diagonal of the angles $\phi_i$.
This resulted in a remarkable "memory" of the original problem
since all the singularities (\ref{singNickel}), forwarded by
Nickel~\cite{nickel-99,nickel-00}, are reproduced, even
the quadratic roots of $1+3w+4w^2=0$ found~\cite{ze-bo-ha-ma-04,ze-bo-ha-ma-05}
  for the linear ODE of $\chi^{(3)}$.

The following steps will be to incorporate, gradually, the
various factors appearing in (\ref{chi3tild}).
One may, for instance, continue to discard the factor $G^{(n)}$,
but incorporate the product on $y_j$ to the 
$\Phi^{(n)}_D(w)$.

In this move towards the $\, n$-particle contributions $\, \chi^{(n)}$
of the Ising model susceptibility, it is natural
to consider the following family of integrals
\begin{eqnarray}
\label{In}
\Phi_H^{(n)} \, \,= \,\,\, {\frac{1}{n!}}  \cdot 
\Bigl( \prod_{j=1}^{n-1}\int_0^{2\pi} {\frac{d\phi_j}{2\pi}} \Bigr)  
\Bigl( \prod_{j=1}^{n} y_j \Bigr)  \cdot  
 {\frac{1\,+\prod_{i=1}^{n}\, 
x_i}{1\,-\prod_{i=1}^{n}\, x_i}} 
\end{eqnarray}
which amounts to getting rid of the $\, (G^{(n)})^2$ in 
(\ref{chi3tild}).

One may restrict this multidimensional integral to a 
simple integral on one angular variable in the same way 
we did it for our previous ``diagonal toy model''
in Section (\ref{Diag}):
\begin{eqnarray}
\label{chinaked2}
 I_{D}^{(n)}\,&=&\,\,\,\,  {{1} \over {n!}} \, \int_0^{2\pi} 
{\frac{d\phi}{2\pi}} \times   \\
&& \, \,\,\,\, 
y^{n-1}(\phi)  \cdot  y ((n-1)\phi) \cdot 
 \,   {\frac{1\, + \,x^{n-1}(\phi)  \cdot  
x ((n-1)\phi)}{1\, -x^{n-1}(\phi)  \cdot  x ((n-1)\phi)}} 
\nonumber 
\end{eqnarray}
For this last family, one obtains for $n=3$, the singularities
found for the linear ODE of the diagonal model integral $\, \Phi^{(3)}_D$ 
of Section (\ref{Diag}) and also the singularities
$(1+3w)(1-5w)=0$. These singularities emerge
as a consequence of inclusion of the $\, y_i$.
Recall that these last singularities do not occur \cite{ze-bo-ha-ma-04} 
in the linear ODE corresponding to $\chi^{(3)}$.
Do they disappear in $\, \chi^{(3)}$ as a consequence of the enlargment 
of multidimensional integration domain, or as a consequence 
of the inclusion of the $\, G^{(n)}$ factor ?

Note that, as we move closer to the original integrand 
and integral (\ref{chi3tild}), for instance considering 
 the multidimensional integral (\ref{In}), 
 the calculations~\cite{ze-bo-ha-ma-04,ze-bo-ha-ma-05,ze-bo-ha-ma-05b} 
(series expansions, search of 
the linear ODE) become much harder. Nevertheless, an exact knowledge
of the singularities of the linear ODE's can be achieved using a brand 
new strategy that amounts to getting very large series for these integrals {\em 
modulo primes}, and in a second step, get the corresponding  linear ODE's  {\em also 
modulo primes}\footnote[5]{This can be seen as a {\em new method for obtaining exact results 
in lattice statistical mechanics}, and, in any case, extremely large 
series expansions. In forthcoming publications with A. J. Guttmann, 
I.Jensen and B. Nickel we will detail such ``extreme'' calculations of
 lattice statistical mechanics {\em modulo the primes} providing more than 1600
coefficients of the high and low temperature
series expansions of the susceptibility $\, \chi$ of the square Ising model, as 
well as more than 2000 coefficients for $\, \chi^{(5)}$, and, {\em for given 
primes}, with 6000 coefficients.}.
Ideally, for a large enough set of such calculations modulo different 
primes, one can get (from a Chinese remainder procedure) the
 exact linear ODE we are seeking for.
In the case where the number of primes and calculations modulo these 
primes are not numerous enough to build the linear ODE, they are 
sufficient to be sure that a certain polynomial (like the ones we displayed in 
Section (\ref{Diag})) actually occurs in the head polynomial of the linear ODE, or discard 
its  occurrence.

\section{Conclusion}
\label{concl}

We have designed two families of ``Ising class integrals''
($Y^{(n)}(w)$, $\Phi^{(n)}_D(w)$), for which we obtained the Fuchsian
linear differential equations for large enough values of $n$.
Each Fuchsian linear differential equation provides a set of singularities.
We have used a direct
resolution (for the first family of integrals)
and our connection matrix method \cite{ze-bo-ha-ma-05c}
(for the second family of integrals) 
to obtain the subset
of singularities occurring in the integrals.
For both families of integrals, we solved the Landau conditions and 
found that the singularities we obtained may identify, or 
{\em even extend}, 
the singularities of the linear  ODE or the integrals.
While it is obvious that the linear ODE, which has this holonomic integral as 
solution, gives all other solutions with their singular behavior, 
it is remarkable, as far as the locus of singularities is concerned,
that the Landau conditions calculations on the integrand give
informations on the other singularities not carried by the integral.

For these ``Ising class integrals'', we found that each family of
integrals is singular in the domain ${\cal W}_c$.
Our integrals are defined for small values of the variable
$w$ and elsewhere by analytical continuation.
This variable $w=s/2/(1+s^2)$ behaves in an equal footing for small and
large values of the variable $s$.
Switching to this last $s$ variable,
we showed that the analytical continuations (from small values of $s$ for which
the integrals are defined) of these
 two families of integrals are singular on points
occurring on the unit circle $\vert s \vert=1$, {\em but also}
 on some points in the
exterior of this circle ($\vert s \vert > 1$). 

In view of these results on our toy integrals, 
we revisited the singular behavior of the third contribution
to the magnetic susceptibility $\tilde{\chi}^{(3)}$ of the Ising model.
We also found here that $\tilde{\chi}^{(3)}$, defined
 as series expansions for small
values of the variable $s$ and analytically continued to
large values of $s$, is {\em actually singular} outside 
the unit circle, $\vert s \vert>1$, 
at the points $2+s+s^2=\, 0$. 
If a similar scenario remains valid for the other  $\chi^{(n)}(s)$,
for $n$ odd, one may imagine that the singularities
of the $\chi^{(n)}(s)$, for $n$ odd, defined in the whole complex
plane of the variable $s$ by analytical continuation
of high temperature $s-$series (valid for $\vert s \vert <1$),
actually correspond to the ``nickellian singularities'' (\ref{Nick}), 
{\em together with a similar "cloud" of singularities} as in Fig. 1
{\em outside} the unit circle ($\vert s \vert >1$).

Our two families of integrals are introduced to mimic the 
$n-$particle contribution $\, \chi^{(n)}$ to 
the susceptibility of the square Ising model. 
In a forthcoming publication we will see to what extend
these two families of integrals, and similar toy models, 
give information on  the singular behavior of these integrals
in the complex plane of the variable $s$ and on the
 singularities that might occur in the
$\, \chi^{(n)}$.

\vskip 0.2cm

\vskip 0.2cm

{\bf Acknowledgments:} 
This work is partially supported by a PICS/CRNS grant.
One of us (S.H) acknowledges the kind hospitality at
the LPTMC where part of this work has been
completed.

\vskip 0.2cm

\section{Appendix A}
\label{appendixB}
We give in this Appendix the order six differential equation 
corresponding
to
\begin{eqnarray}
{\tilde Y}^{(3)}(w)\,\,  =\,\, \,  \,  \, 
 \int_0^{2\pi }\frac{d\phi _{1}}{2\pi }\, \int_{0}^{2\pi }
\frac{d\phi _{2}}{2\pi } \cdot 
 y_1 \, y_2 \, y_3  
\end{eqnarray}
that reads ($F(w)={\tilde Y}^{(3)}(w)$)
\begin{eqnarray}
\sum_{n=0}^{6}\,a_{n}(w)\cdot 
{\frac{{d^{n}}}{{dw^{n}}}}F(w)\,\,=\,\,\,\,\,0
\end{eqnarray}
where
\begin{eqnarray}
a_6 (w) & = &
\left(1-5\,w \right)  \left( 1+3\,w \right)  
\left( 1+3\,w+4\,{w}^{2} \right)  \left( 1+4\,w \right)^{2} 
\left(1-4\,w \right)^{4}  
\nonumber \\
&&\qquad   \left(1-w \right)\, {w}^{6} \cdot P_6(w)  \nonumber \\
a_5 (w) & = &
2\, \left( 1+4\,w \right)  \left( -1+4\,w \right)^{3}\,{w}^{5} \cdot  
P_5(w) \nonumber \\
a_4 (w)  &=&
2\, (1-4\,w)^{2}\,{w}^{4} \cdot P_4(w), \quad
a_3(w) =\, -4\,  (1-4\,w)\cdot {w}^{3}  \cdot P_3(w) \nonumber 
\\
a_2(w) &=&4\,{w}^{2} \cdot P_2(w), \quad \quad
a_1(w) = \, 8\,{w} \cdot  P_1(w), \quad \quad
a_0(w) = \, P_0(w) \nonumber 
\end{eqnarray}
with
\begin{eqnarray}
&& P_6(w)=\, 464889446400\,{w}^{17}-5967097036800\,{w}^{16}+2518865879040\,{w}^{15} \nonumber \\
&& +1205147090944\,{w}^{14}-209927829504\,{w}^{13}-23192809472\,{w}^{12} \nonumber \\
&& -94576000000\,{w}^{11}+15333496352\,{w}^{10}
+6474551520\,{w}^{9}+434664056\,{w}^{8}  \nonumber \\
&& -768760124\,{w}^{7}+47392381\,{w}^{6}+24032176\,{w}^{5}-3740777\,{w}^{4}  \nonumber \\
&& +106680\,{w}^{3}+7443\,{w}^{2}-364\,w+49 \nonumber \\
&& P_5(w)=\,  5020806021120000\,{w}^{24}-67498529469235200\,{w}^{23} \nonumber \\
&& +30557536675430400\,{w}^{22}+38088998121308160\,{w}^{21} \nonumber \\
&& -4237195746729984\,{w}^{20}-9069732118712320\,{w}^{19}-2846232832169984\,{w}^{18} \nonumber \\
&& +241117241392640\,{w}^{17}+618725352473088\,{w}^{16}
+201166769452896\,{w}^{15} \nonumber \\
&& -61245521268224\,{w}^{14}-22970672645976\,{w}^{13}+1946216478788\,{w}^{12} \nonumber \\
&& +1256316716689\,{w}^{11}-2295775019\,{w}^{10}-38685730792\,{w}^{9} \nonumber \\
&& +7182763949\,{w}^{8}-2662652062\,{w}^{7}+133819572\,{w}^{6}+107083800\,{w}^{5} \nonumber \\
&& -16853044\,{w}^{4} +499141\,{w}^{3}+34119\,{w}^{2}-784\,w+147 \nonumber \\
&& P_4(w)=\,  562330274365440000\,{w}^{26}-7836264826719436800\,{w}^{25} \nonumber \\
&& +3202063539476889600\,{w}^{24}+3620634209723351040\,{w}^{23} \nonumber \\
&& -548610793223356416\,{w}^{22}-626862114022522880\,{w}^{21} \nonumber \\
&& -294517075356614656\,{w}^{20}+30188250588135424\,{w}^{19} \nonumber \\
&& +71955514644985856\,{w}^{18}+6645859876651520\,{w}^{17} \nonumber \\
&& -7980786065196160\,{w}^{16}-956452174732544\,{w}^{15} \nonumber \\
&& +546200976422432\,{w}^{14}+127674138396992\,{w}^{13} \nonumber \\
&&  -45199723462020\,{w}^{12}-9028528613608\,{w}^{11}+2584065207968\,{w}^{10} \nonumber \\
&& +428924564347\,{w}^{9}-59295462952\,{w}^{8}-29452365937\,{w}^{7} \nonumber \\
&& +3600235371\,{w}^{6}+631583771\,{w}^{5}-118172923\,{w}^{4}+3790389\,{w}^{3} \nonumber \\
&& +186393\,{w}^{2}-5138\,w+931 \nonumber \\
&&P_3(w)=\, 2811651371827200000\,{w}^{27}-41266384607379456000\,{w}^{26} \nonumber \\
&& +24773736838201344000\,{w}^{25}+11079955148990054400\,{w}^{24} \nonumber \\
&& -4824757963497799680\,{w}^{23}-848931834754990080\,{w}^{22} \nonumber \\
&& -1254549778859409408\,{w}^{21}+332436165839831040\,{w}^{20} \nonumber \\
&& +230844285233797120\,{w}^{19}+23397830675772928\,{w}^{18} \nonumber \\
&& -35731466794253184\,{w}^{17}-7015978530910848\,{w}^{16} \nonumber \\
&& +3334508426832544\,{w}^{15}+129769752723552\,{w}^{14} \nonumber \\
&& +64823043832572\,{w}^{13}+9687172326860\,{w}^{12}-27269045355036\,{w}^{11} \nonumber \\
&& +668812543530\,{w}^{10}+1232145138001\,{w}^{9}+217652915670\,{w}^{8} \nonumber \\
&& -113666172030\,{w}^{7}+2636314376\,{w}^{6}+2843958768\,{w}^{5} \nonumber \\
&& -343848072\,{w}^{4}+7774186\,{w}^{3}+533670\,{w}^{2}-21133\,w+2254 \nonumber \\
&&P_2(w)=\, 8434954115481600000\,{w}^{28}-130054335243485184000\,{w}^{27} \nonumber  \\
&& +101906251242799104000\,{w}^{26}+14913046820133273600\,{w}^{25} \nonumber \\
&& -16311072130508390400\,{w}^{24}+3172467314793676800\,{w}^{23} \nonumber \\
&& -7523898588462022656\,{w}^{22}+4342034847529598976\,{w}^{21} \nonumber \\
&& +349567990475905024\,{w}^{20}-1019978058716864000\,{w}^{19} \nonumber \\
&& +135175191464703616\,{w}^{18}+154325934113136384\,{w}^{17} \nonumber \\
&& -39182557846936736\,{w}^{16}-6064402640954016\,{w}^{15} \nonumber \\
&& -721601939503388\,{w}^{14}+778716878219200\,{w}^{13}+414373001033428\,{w}^{12} \nonumber \\
&& -145383247830248\,{w}^{11}-7935890967120\,{w}^{10}+857310545482\,{w}^{9} \nonumber \\
&& +2441550352242\,{w}^{8}-452151200437\,{w}^{7}-32488744995\,{w}^{6} \nonumber \\
&& +15597930449\,{w}^{5}-1389870493\,{w}^{4}+21276545\,{w}^{3} \nonumber \\
&& +2254771\,{w}^{2}-106015\,w+7987 \nonumber \\
&&P_1(w)=\, 421747705774080000\,{w}^{28}-6710038906798080000\,{w}^{27} \nonumber \\
&& +4873843971588096000\,{w}^{26}+1036215739573862400\,{w}^{25} \nonumber \\
&& -616703179652628480\,{w}^{24}-120979270258114560\,{w}^{23} \nonumber \\
&& +364788984254619648\,{w}^{22}-1023471173198333952\,{w}^{21} \nonumber \\
&& -201098747980344320\,{w}^{20}+953808829975566464\,{w}^{19} \nonumber \\
&& -246857972154239232\,{w}^{18}-152394614023808544\,{w}^{17} \nonumber \\
&& +55269675209779600\,{w}^{16}+3979478375375692\,{w}^{15} \nonumber \\
&& +612261498482876\,{w}^{14}-759726033790536\,{w}^{13}-527516229973280\,{w}^{12} \nonumber \\
&& +188404041710426\,{w}^{11}+5343977363342\,{w}^{10}-599703686985\,{w}^{9} \nonumber \\
&& -2917831385349\,{w}^{8}+538227200312\,{w}^{7}+39103736504\,{w}^{6} \nonumber \\
&& -18721639514\,{w}^{5}+1671294354\,{w}^{4}-25882894\,{w}^{3} \nonumber \\
&& -2676882\,{w}^{2}+122171\,w-9261  \nonumber \\
&&P_0(w)=\, -32133158535168000\,{w}^{26}+695978420915404800\,{w}^{25} \nonumber \\
&& -544662214410240000\,{w}^{24}+3609286260943749120\,{w}^{23} \nonumber \\
&& -7536908069936824320\,{w}^{22}+10191253169267146752\,{w}^{21} \nonumber \\
&& +2316960321752137728\,{w}^{20}-8808777472942522368\,{w}^{19} \nonumber \\
&& +2190279912630165504\,{w}^{18}+1472339197313387520\,{w}^{17} \nonumber \\
&& -535796399651893248\,{w}^{16}-40698424759766784\,{w}^{15} \nonumber \\
&& -1761836408896896\,{w}^{14}+6557432066372256\,{w}^{13} \nonumber \\
&& +5053923219182592\,{w}^{12}-1884372068029344\,{w}^{11}-14699422810656\,{w}^{10} \nonumber \\
&& +1982678836896\,{w}^{9}+27382956586968\,{w}^{8}-5031984639480\,{w}^{7} \nonumber \\
&& -368590752120\,{w}^{6}+176119304760\,{w}^{5}-15746069112\,{w}^{4} \nonumber \\
&& +246121272\,{w}^{3}+24962808\,{w}^{2}-1111320\,w+84672  \nonumber
\end{eqnarray}

\section{Appendix B}
\label{appendixC}
In this Appendix, we derive the $Y^{(n)}$ expansion (\ref{FN 
expansion}).
At the first step, the $n-1$ dimensional integral is translated to a
$n$ dimensional one, the angular constraint (\ref{they}) being taken
into account by a Dirac delta distribution $\delta (\sum_{i=1}^{n}\phi 
_{i})$
which is Fourier expanded as  (with $Z_{p}=\exp (i\, \phi _{p})$) 
\begin{eqnarray}
\label{delta Fourier exp}
2\pi \cdot  \delta (\sum_{i=1}^{n}\phi _{i})\, =\, \, 
\sum_{k=-\infty }^{\infty}\Bigl( Z_{1}\, Z_{2}\, \cdots\, Z_{n} \Bigr)^{k}  
\end{eqnarray}
The integrals (\ref{defFN}) read
\begin{eqnarray}
&&Y^{(n)}(w) =\,  \sum_{k=-\infty }^{\infty }\,
\int_{0}^{2\pi }\frac{d\phi _{1}}{2\pi } \int_{0}^{2 \pi }\frac{d\phi _{2}}{2\pi }\,  \cdots \, 
\int_{0}^{2\pi }\frac{d\phi _{n}}{2\pi } \cdot 
\Bigl(\prod_{i=1}^{n} y_{i}^{2}\cdot  Z_{i}^{k} \Bigr) \nonumber \\
&&\qquad \qquad =\,  \sum_{k=-\infty }^{\infty }\,  \Bigl(  
\int_{0}^{2\pi } \frac{d\phi}{2\pi }
\, y^{2}\cdot  Z^{k}  \Bigr)^{n} 
\end{eqnarray}
From the Fourier expansion (\ref{FourExp Y2}), one has the trivial
integration rule
\begin{eqnarray}
\int_{0}^{2\pi } \frac{d\phi }{2\pi }\left( y^{2}\, Z^{k}\right) \, =\,  \,
\, 4\, w^{2}\cdot  D(k) 
\end{eqnarray}
One  gets 
\begin{eqnarray}
Y^{(n)}(w)\, \, =\, \, \, (4w^{2})^{n} \cdot \sum_{k=-\infty }^{\infty }\, 
D(k)^{n} 
\end{eqnarray}
and with the definition (\ref{defD(k)}), we
finally obtain (\ref{FN expansion}).

\section{Appendix C}
\label{appendixD}
In this Appendix we give for the $y^2-$product model, the differential 
equation
for $n=4$ and the polynomials $R_1$ and $R_2$ corresponding
to the solution $\, Y^{(5)}(w)$.
The differential equation for $Y^{(4)}(w)$ reads
\begin{eqnarray}
a_2 \, {\frac{d^2}{dw^2}} Y^{(4)}\,  + a_1 \, {\frac{d}{dw}} Y^{(4)}\, 
 +a_0 \,  Y^{(4)}\,\,  =\,\,\,  0
\end{eqnarray}
where
\begin{eqnarray}
&&a_2 \, =\, 
 \left( 1-16\,{w}^{2} \right) ^{2} \left( 1-4\,{w}^{2} \right) 
 \left( 1-20\,{w}^{2}+16\,{w}^{4}-16\,{w}^{6} \right)
\, {w}^{2}  \cdot  P_2(w), \nonumber \\
&&a_1 \, =\,  2\, w \left( 1-16\,{w}^{2} \right) \, P_1(w), \qquad a_0=8\, P_0(w) \nonumber
\end{eqnarray}
with:
\begin{eqnarray}
&&P_2 \, =\,   -15+1086\,{w}^{2}-28106\,{w}^{4}+328716\,{w}^{6}-1946216\,{w}^{8} \nonumber \\
&& +4791440\,{w}^{10}-4697088\,{w}^{12}+8682368\,{w}^{14}-17308416\,{w}^{16}  \nonumber \\
&& +1781760\,{w}^{18},   \nonumber \\
&&P_1 \, =\,   75-7716\,{w}^{2}+319862\,{w}^{4}-7044848\,{w}^{6}+92471304\,{w}^{8} \nonumber \\
&& -761556288\,{w}^{10}+3861020800 \,{w}^{12}-11558422912\,{w}^{14} \nonumber \\
&& +22069511424\,{w}^{16}-42440728576\,{w}^{18}+101426089984\,{w}^{20} 
\nonumber \\
&& -147848548352\,{w}^{22}+72672886784\,{w}^{24} -17528389632\,{w}^{26} \nonumber \\
&& +2736783360\,{w}^{28},  \nonumber \\
&&P_0 \, =\,  -45+5400\,{w}^{2}-261002\,{w}^{4}+6787240\,{w}^{6}-107032250\,{w}^{8} \nonumber \\
&& +1079592480\,{w}^{10}-6995904504\,{w}^{12} +28478625024\,{w}^{14} \nonumber \\
&& -72563446848\,{w}^{16}+142196898048\,{w}^{18} -402528466944\,{w}^{20}  \nonumber \\
&& +1067417646080\,{w}^{22}-1296486492160\,{w}^{24} +418540593152\,{w}^{26} \nonumber \\
&& -55343579136\,{w}^{28}+13683916800\,{w}^{30}.  \nonumber
\end{eqnarray}

For the solution $\, Y^{(5)}(w)$, the polynomial $P^{(5)}_1$ and 
$P^{(5)}_2$
are given in (\ref{sets5}) and
\begin{eqnarray}
&&R^{(5)}_1 \, =\,  15-950\,{w}^{2}-550\,{w}^{3}+25581\,{w}^{4}
 +14856\,{w}^{5}-363086\,{w}^{6}  \nonumber \\
&& \quad  -57174\,{w}^{7}+2727087\,{w}^{8}-1243960\,{w}^{9}-9109540\,{w}^{10} \nonumber \\
&& \quad +10083482\,{w}^{11}+8271200\,{w}^{12} -23345456\,{w}^{13}-2557976\,{w}^{14} \nonumber \\
&& \quad +19604672\,{w}^{15}+3295040\,{w}^{16}-1531392\,{w}^{17}-513024\,{w}^{18},  \nonumber \\
&&R^{(5)}_2 \, =\,  3+6\,w-103\,{w}^{2}-292\,{w}^{3}
 +1207\,{w}^{4}+2214\,{w}^{5}-7360\,{w}^{6}  \nonumber \\
&& \quad +3674\,{w}^{7}+6896\,{w}^{8}-21328\,{w}^{9}
-13080\,{w}^{10}-96\,{w}^{11}-256\,{w}^{12}.  \nonumber 
\end{eqnarray}

\section{Appendix D}
\label{appendixA}
The differential equation satisfied by $\Phi_D^{(3)}$ reads:
\begin{eqnarray}
\sum_{n=1}^{4}\,a_{n}(w)\cdot 
{\frac{{d^{n}}}{{dw^{n}}}}F(w)\,\,=\,\,\,\,\,0
\end{eqnarray}
where
\begin{eqnarray}
a_4 (w) & = &
 \left( 1-w \right) \left( 1+2w \right)  \left( 1-4w
 \right)  \left( 1+4w \right)  \left( 1+3w+4{w}^{2} \right)\, {w}^{2}
 \cdot  P_4(w), \nonumber \\
a_3 (w) & = & w \cdot  P_3(w), \qquad
a_2 (w)  =  P_2(w), \qquad
a_1(w) = P_1(w)  \nonumber
\end{eqnarray}
with
\begin{eqnarray}
&&P_4(w)\, =\,   -18-126\,w-536\,{w}^{2}-581\,{w}^{3}+11332\,{w}^{4}+56216\,{w}^{5} 
\nonumber \\
&&  +141103\,{w}^{6}  +316146\,{w}^{7}+324516\,{w}^{8}
-102512\,{w}^{9} +512\,{w}^{10} \nonumber \\
&&  +104448\,{w}^{11},  \nonumber \\
&&P_3(w) =\,  -36-450\,w-270\,{w}^{2}+17199\,{w}^{3}
+110892\,{w}^{4} +331122 \,{w}^{5}  \nonumber \\
&& -365893\,{w}^{6}  -9173304\,{w}^{7}-40917443\,{w}^{8}-92069955\,{w}^{9}  \nonumber \\
&& -128675122\,{w}^{10}-89628548\,{w}^{11}  +223226064\,{w}^{12} \nonumber \\
&& +725436224\,{w}^{13}+509586688\,{w}^{14}-185729024\,{w}^{15} \nonumber \\
&& +25788416\,{w}^{16}+133693440\,{w}^{17},  \nonumber \\
&&P_2(w)\, =\, 36+396\,w+6114\,{w}^{2}+59550\,{w}^{3}
+302337\,{w}^{4}+857367\,{w}^{5} \nonumber \\
&& +887406\,{w}^{6}  -6754260\,{w}^{7}-39314109\,{w}^{8}-48807357\,{w}^{9} \nonumber \\
&& +175442484\,{w}^{10}+683394108\,{w}^{11}  +1327431600\,{w}^{12} \nonumber \\
&& +2198750784\,{w}^{13}+1607543040\,{w}^{14}-479219712\,{w}^{15} \nonumber \\
&& +35880960\,{w}^{16}+320864256\,{w}^{17}, \nonumber \\
&&P_1(w) \,=\, 108+1512\,w+24282\,{w}^{2}+230094\,{w}^{3}+1072500\,{w}^{4} \nonumber \\
&& +3034020\,{w}^{5}+6713730\,{w}^{6} +7870578\,{w}^{7}+26737134\,{w}^{8} \nonumber \\
&& +253594938\,{w}^{9}+799267644\,{w}^{10}+1253301264\,{w}^{11} \nonumber \\
&& +1490308224\,{w}^{12} +1015581696\,{w}^{13}-338632704\,{w}^{14} \nonumber \\
&& -1179648\,{w}^{15}  +160432128\,{w}^{16}. \nonumber
\end{eqnarray}

The differential equation satisfied by $\Phi_D^{(4)}$ reads:
\begin{eqnarray}
\sum_{n=1}^{4}\, a_{n}(w)\cdot 
{\frac{{d^{n}}}{{dw^{n}}}}F(w)\,\,=\,\,\,\,\,0
\end{eqnarray}
where
\begin{eqnarray}
&&a_4 (w) \,  = \,{w}^{3} \, (1+4\,w)  \, (1+2\,w)  \, (1-2\,w)
  \, (1- 4\,w) \cdot P_4(w), \nonumber \\
&&a_3 (w) \, = \,2\,w^4 \cdot P_3(w), \quad
a_2 (w)  = 3 w \cdot P_2(w), \quad a_1(w) = 3 \cdot P_1(w)  \nonumber
\end{eqnarray}
with
\begin{eqnarray}
&&P_4(w) \,= \, 1 +19\,{w}^{2} +314\,{w}^{4} + 512\,{w}^{6}, \nonumber \\
&&P_3(w) \, = \,-81-1010\,{w}^{2}-1756\,{w}^{4}+89408\,{w}^{6}+114688\,{w}^{8},\nonumber \\
&&P_2(w) \, =\, -1-53\,{w}^{2}+98304\,{w}^{10}
+117888\,{w}^{8}+742\,{w}^{4}+24896\,{w}^{6},  \nonumber \\
&&P_1(w) \, =\,   1+53\,{w}^{2}+ 2042\,{w}^{4}+15616\,{w}^{6}. 
 \nonumber
\end{eqnarray}

\section{Appendix E}
\label{B}
In this Appendix, we give the proof for the solutions of the Landau
conditions. One may use the form of the integrand given in 
(\ref{phinratio}),
we instead use another form to show clearly how the singularities 
of the "Galois" companions are reproduced.

The variable $x_j$ given in (\ref{thex}) from the definition
(\ref{defangzeta}) can be written as:
\begin{eqnarray}
\label{xjzeta}
x_j(\phi_j)\, = \, \exp \left( i\, \zeta_j \right)
\end{eqnarray}
The integral (\ref{chinaked}) then becomes
\begin{eqnarray}
\label{defGN2}
\Phi_{D}^{(n)}\,= \,\,\, {\frac{1}{n!}}\, \int_{0}^{2\pi }
 \frac{d\phi }{2\pi }\, \frac{i}{\tan (\psi /2)},
\qquad \psi =\,(n-1) \zeta \,+\zeta_n
\end{eqnarray}
for which, the Landau conditions read:
\begin{eqnarray}
&& \alpha_1 \cdot  \psi \, =\,\,0, \qquad \quad 
\quad  \alpha_2 \cdot \phi =\,\,0,  \nonumber \\
&& \alpha_3 \cdot  \sin \left(\zeta \right)^2\, =\,\,0, \qquad \quad 
\quad  \alpha_4 \cdot \sin \left(\zeta_n \right)^2\, =\,\,0, \nonumber 
\\
&& \alpha_1 \cdot  \psi^{'} \, +\, 
\alpha_2  \, +\, 2 \alpha_3 \cdot \sin(\phi)\cos(\zeta) \,  \\
&& \qquad \qquad \qquad +\,2 \alpha_4 \cdot 
\sin((n-1)\phi)\cos(\zeta_n)
 \,\,  =\,\,\, 0 \nonumber
\nonumber
\end{eqnarray}
where the parameters $\alpha_i$, $i=1,\cdots 4$, should not be, all, 
equal to 
zero. In the sequel, we give the proofs along the possible values (zero 
or not)
 of the two parameters $\alpha_3$ and $\alpha_4$.

\subsection{$\alpha_3= \alpha_4 =0$}

In this configuration, we are discarding the conditions on the branch 
points.
The Landau conditions for $\alpha_1$ and $\alpha_2$ not both
equal to zero give:
\begin{eqnarray}
\label{Landaueq1GNbis}
\zeta _{n} +(n-1)\zeta \, =\,\,  0,
 \quad \quad \quad \quad \quad  \phi\, =\,\, 0
\end{eqnarray}
or:
\begin{eqnarray}
\label{Landaueq2GNbis}
\zeta _{n} + (n-1)\zeta \, =\,\,  0, 
 \quad  \quad \quad 
\frac{\sin ((n-1)\phi )}{\sin(\zeta_{n}) }\,\, =\, \frac{\sin (\phi )}{\sin (\zeta)}
\end{eqnarray}
These equations correspond (with $\sin(\zeta) \ne 0$),
respectively, to an {\em end-point}
singularity and to  a {\em pinch} singularity.
The condition (\ref{Landaueq1GNbis}) being contained in 
(\ref{Landaueq2GNbis})
for $\phi=0$, one may consider the last.

The condition $\, \, \zeta _{n} + (n-1)\zeta =\, 0\,\,  $ is {\em equivalent} to 
\begin{eqnarray}
\label{eqcos}
&& \cos(\zeta_n) \,  = \, \,  \cos\left( (n-1)\zeta \right), \\
\label{eqsin}
&& \sin(\zeta_n) \,  = \,\,   -\sin\left( (n-1)\zeta \right).
\end{eqnarray}
Using (\ref{defangzeta}) for $\cos(\zeta_n)$ written as
\begin{eqnarray}
 \cos(\zeta_n) \,  =  \,\,   {1 \over 2w} -\cos((n-1) \phi) \,\,    =  \,  
\,   \, {{1 }
\over {2 w}} \,  -T_{n-1}\left(\cos(\phi) \right)
\end{eqnarray}
one obtains for (\ref{eqcos}):
\begin{eqnarray}
 {1 \over 2w} -T_{n-1}(\cos(\phi))\,   \,  = \,  \, \,  
\cos \Bigl( (n-1) \cos^{-1} \left( {1 \over 2w} -\cos(\phi) \right) 
\Bigr)
\end{eqnarray}
Denoting $v\, = \, \cos(\phi)$, and by definition of the Chebyshev 
polynomials, 
one gets
\begin{eqnarray}
\label{eq2}
  {1 \over 2w} -T_{n-1}(v) \, \,    = \,  \,  \,  
T_{n-1} \Bigl(  {1 \over 2w} -v \Bigr)
\end{eqnarray}
which is the first condition given in (\ref{newsetL3}).
One has
\begin{eqnarray}
 \sin\left( (n-1) \zeta \right)\, \,   =\, \, \,  
 \sqrt{1-\left({1 \over 2w}-\cos(\phi) \right)^2 } \cdot  U_{n-2} 
\left({1 \over 2w}-\cos(\phi) \right) \nonumber 
\end{eqnarray}
and taking $\,\sin(\zeta_n)$ as
\begin{eqnarray}
\label{defsin}
 \sin(\zeta_n)\, = \, \, \sqrt{1-\cos(\zeta_n)^2}=
  \sqrt{1-\Bigl( {1 \over 2w}- T_{n-1}\left(\cos(\phi) \right)   
\Bigr)^2}
\end{eqnarray}
the condition (\ref{eqsin}) gives
\begin{eqnarray}
\label{setL3cons}
&&\sqrt{1-\Bigl( {1 \over 2w}-T_{n-1}(v) \Bigr)^2}\, \\
&&\qquad \qquad +\sqrt{1-\Bigl( {1 \over 2w}-v \Bigr)^2} \, \cdot \,   
U_{n-2} \left( {1 \over 2w}-v\right)\,=\,0 \nonumber 
\end{eqnarray}
which is the third condition given in (\ref{newsetL3})
and is a constraint to (\ref{eq2}).

The second condition in (\ref{Landaueq2GNbis}) is automatically 
satisfied for $\phi=0$ and 
$\phi=\pi$.
Plugging these values in (\ref{eq2}) and (\ref{setL3cons}) gives
respectively
\begin{eqnarray}
\label{setL1cons}
 1 \,\, +  U_{n-2} \left({1 \over 2w}-1 \right)\,=\,\,0, 
\end{eqnarray}
\begin{eqnarray}
\label{setL2cons}
\sqrt{1-\left( {1 \over 2w}-(-1)^{n-1} \right)^2 } \,  + \, 
\sqrt{1-\left( {1 \over 2w}+1 \right)^2 }\cdot
 U_{n-2}\left( {1 \over 2w}+1 \right)\, =\, 0 \nonumber 
\end{eqnarray}
which are the constraints given in (\ref{newsetL1}) and 
(\ref{newsetL2}).
The last have solutions only for $\, n$ even.

At this point, the second condition in (\ref{Landaueq2GNbis})
has not been used.
Using (\ref{eqsin}) in this last condition, one obtains
\begin{eqnarray}
{\frac{\sin((n-1) \left( \cos^{-1}(v) \right) }{\sin\left( (n-1) 
\cos^{-1}(1/2w-z) \right) }} \, +
\, {\frac{\sin\left(\cos^{-1}(v) \right)}{\sin \left(\cos^{-1}(1/2w-v) 
\right)}} \,=\,\, 0
\end{eqnarray}
yielding:
\begin{eqnarray}
U_{n-2}(v) \, \, - U_{n-2} \left( {1 \over 2w} -v \right)\,\, =\,\, 0
\end{eqnarray}
which is the second of equation in  (\ref{newsetL3}).

Let us sum up for this case. Using the first condition in 
(\ref{Landaueq2GNbis}),
 giving {\em both} conditions (\ref{eqcos}) and (\ref{eqsin}), leads
to a smaller set of singularities. While imposing the only condition 
(\ref{eqcos})
allows the appearance of the singularities of the "Galois companions".

\subsection{$\alpha_3 \ne  0$}

In this configuration, if $\alpha_4 =0$, the condition 
$\,\sin \left(\zeta \right)=\,0\,$ gives directly $w= \pm 1/4$.

For $\alpha_4 \ne 0$, we have 
$\sin \left(\zeta \right)=\sin \left(\zeta_n \right)=0$, which, using 
the definition of $\cos \left(\zeta \right)$ and $\cos \left(\zeta_n 
\right)$
leads to 
\begin{eqnarray}
 T_{n-1} \left( {\frac{1}{2w}}\, -(-1)^k \right)\,\, + (-1)^m\, \,
-{\frac{1}{2w}}\,\, =\,\, 0, 
\end{eqnarray}
where $k$ and $m$ are integers.

\subsection{$\alpha_3 = 0$ and $\alpha_4 \ne 0$}
In this configuration, additional polynomials are obtained for 
$\alpha_1 \ne 0$ resulting in
\begin{eqnarray}
(n-1) \zeta \,+\zeta_n \,=\,0, \qquad   \hbox{with,}  \qquad \cos 
\left(\zeta \right)^2 \,=\,1
\end{eqnarray}
leading to
\begin{eqnarray}
&& T_{n-1}(v)\, - \left( {\frac{1}{2w}}\, \pm 1 \right) \, =\,\, 0, \quad 
\quad \hbox{and} \quad \quad 
 T_{n-1}\left( {\frac{1}{2w}}\, -v \right)\,\, \pm 1 \, = \,\, 0 \nonumber 
\end{eqnarray}
which is solved by elimination of $v$.

\section{Appendix F}

The order one and order two differential operators occurring in the
order-four linear differential operator 
\begin{eqnarray}
Ds \oplus {\cal L}_3(s)\, \,  = \,\, \, 
 Ds \oplus \Bigl({\cal L}_2(s) \cdot  {\cal L}_1(s) \Bigr)
\end{eqnarray}
written in the variable $\, s$ and 
corresponding to $\Phi^{(3)}_D$ read: 
\begin{eqnarray}
\label{L1L2}
 {\cal L}_1(s) \, = \, \,  \, \,
 Ds \, -{1 \over 2}\, {Q_0 \over Q_1}, \quad \quad 
 {\cal L}_2(s) \, = \,  \,\,  Ds^2 \, 
 + {P_1 \over P_2}\cdot Ds \,  + {P_0 \over P_2}   \nonumber
\end{eqnarray}
with
\begin{eqnarray}
&& Q_1\, = \, \, \, \left( 1+{s}^{2} \right)  \left( {s}^{2}+s+1 \right) \nonumber  \\
&& \qquad \times \left( 2\,{s}^{2}-s+2 \right)  \left( 2\,{s}^{2}+s+1 \right) 
 \left( {s}^{2}+s+2 \right)  \left( 2\,{s}^{2}+s+2 \right),   \nonumber  \\
&& Q_0\, =\,\,\,
  \left( s^2-1 \right)\nonumber  \\
&& \quad \times
 \left( 8\,{s}^{8}+12\,{s}^{7}+26\,{s}^{6}+19\,{s}^{5}+32\,{s}^{4}+19
\,{s}^{3}+26\,{s}^{2}+12\,s+8 \right) \nonumber 
\end{eqnarray}
and:
\begin{eqnarray}
&& P_2\, =\,\,
s \left( 1+{s}^{2} \right) \left( 2\,{s}^{2}+s+2 \right)^{2}
 \left( s^2-1 \right) ^{2} \left( 2\,{s}^{2}+s+1 \right)
  \left( {s}^{2}+s+2 \right),  \nonumber \\
&& \quad \quad \left( {s}^{2}+s+1 \right) \left( 2\,{s}^{2}-s+2 \right) 
 \Bigl( 64\,{s}^{16}+328\,{s}^{14}+596\,{s}^{13}+4434\,{s}^{12}  \nonumber \\
&& \quad \quad +6621\,{s}^{11}+15377\,{s}^{10} +16897\,{s}^{9}
+22822\,{s}^{8}+16897\,{s}^{7}+15377\,{s}^{6}  \nonumber \\
&& \quad \quad +6621\,{s}^{5}+4434\,{s}^{4}   +596\,{s}^{3}+328\,{s}^{2}+64 \Bigr),  \nonumber \\
&& P_1\, =\,\,  \left( s^2-1 \right)  \left( 2\,{s}^{2}+s+2 \right)  \left( 1+{s}^{2} \right) 
 \Bigl(2048\,{s}^{28}+3840\,{s}^{27}+17792\,{s}^{26}  \nonumber \\
&& \quad \quad +63552\,{s}^{25}+381504\,{s}^{24}+1217888\,{s}^{23}
+3351080\,{s}^{22}+6980836\,{s}^{21} \nonumber \\
&& \quad \quad +12444792\,{s}^{20}+18550693\,{s}^{19}+23711235\,{s}^{18}
+26152837\,{s}^{17}  \nonumber \\
&& \quad \quad +24357824\,{s}^{16}+19695118\,{s}^{15}
+12653972\,{s}^{14}+7020790\,{s}^{13}  \nonumber \\
&& \quad \quad +2315882\,{s}^{12}+789433\,{s}^{11}-161991\,{s}^{10}
+312961\,{s}^{9}+197622\,{s}^{8} \nonumber \\
&& \quad \quad +343348\,{s}^{7}+123248\,{s}^{6}+49472\,{s}^{5}
-28512\,{s}^{4}-28608\,{s}^{3}  \nonumber \\
&& \quad \quad -17152\,{s}^{2}-3840\,s-1024\Bigr),
\nonumber \\
&& P_0 \,  =\, \, 2048\,{s}^{33}+3072\,{s}^{32}+17152\,{s}^{31}
+69760\,{s}^{30}+526400\,{s}^{29} \nonumber \\
&& \quad \quad +1787232\,{s}^{28}+5094288\,{s}^{27}+11072344\,{s}^{26}
+19781548\,{s}^{25} \nonumber \\
&& \quad \quad +27467354\,{s}^{24} +26981054\,{s}^{23}+5123635\,{s}^{22}
-46141730\,{s}^{21} \nonumber \\
&& \quad \quad -134419668\,{s}^{20}  -242023980\,{s}^{19}
-354116927\,{s}^{18}-430640772\,{s}^{17} \nonumber \\
&& \quad \quad -463578366\,{s}^{16} -433931810\,{s}^{15}-371127411\,{s}^{14}
-281456218\,{s}^{13} \nonumber \\
&& \quad \quad -203391640\,{s}^{12} -133972232\,{s}^{11}-89380041\,{s}^{10} \nonumber \\
&& \quad \quad -55343916\,{s}^{9} -34593136\,{s}^{8} -19033912\,{s}^{7}
-9514544\,{s}^{6}-3824384\,{s}^{5}  \nonumber \\
&& \quad \quad -1169344\,{s}^{4} -213760\,{s}^{3}-768\,{s}^{2}+13824\,s+2048. \nonumber
\end{eqnarray}

The polynomials $D_0$, $\, D_K$ and $\, D_E$ occurring in the solutions
(\ref{upto}) of the
order two linear differential operator ${\cal L}_2(s)$ are:
\begin{eqnarray}
&& D_0\, =\,
 s \, (s-1)  \, (1+s)  \, (1+{s}^{2})
 \left( 2+s+2\,{s}^{2} \right)
 \left( 2-s+2\,{s}^{2} \right) \nonumber \\
&& \qquad \quad  \left( 1+s+{s}^{2} \right)
 \left( 2+s+{s}^{2} \right)
 \left( 1+s+2\,{s}^{2} \right),
  \nonumber \\
&& D_K \, =\,
 \, (s-1) \left( 1+s \right) ^{2} \left( 1+{s}^{2} \right)^{2} \nonumber \\
&& \qquad\quad   \left( 8\,{s}^{8}+4\,{s}^{7}+22\,{s}^{6}+9\,{s}^{5}
+93\,{s}^{4}+60\,{s}^{3}+100\,{s}^{2}+32\,s+32 \right),  \nonumber \\
&& D_E \, =\,
2\, \left( 1+{s}^{2} \right) \left( 1+s \right) ^{2}
\left( 2+s+2\,{s}^{2} \right) \nonumber \\
&& \qquad \quad \left( 8\,{s}^{6}-2\,{s}^{5}+17\,{s}^{4}-10\,{s}^{3}+17\,{s}^{2}-2\,s+8 \right).
  \nonumber
\end{eqnarray}


\vskip 0.5cm

\begin{thebibliography}{10}

\bibitem{nickel-99} B. Nickel, 
 J. Phys. A: Math. Gen. \textbf{32} (1999) 3889-3906


\bibitem{nickel-00} B. Nickel,
 J. Phys. A: Math. Gen. \textbf{33} (2000)  1693-1711


\bibitem{wu-mc-tr-ba-76}  T.T. Wu, B.M. McCoy, C.A. Tracy and E. 
Barouch, Phys. Rev. {\bf B 13} (1976) 316-374 

\bibitem{ze-bo-ha-ma-04}
N. Zenine, S. Boukraa, S. Hassani, J.M. Maillard,
J. Phys. A: Math. Gen. {\bf 37} (2004) 9651-9668 and  
arXiv:math-ph/0407060

\bibitem{ze-bo-ha-ma-05}
N. Zenine, S. Boukraa, S. Hassani, J.M. Maillard,
J. Phys. A: Math. Gen. {\bf 38} (2005) 1875-1899
 and arXiv:hep-ph/0411051

\bibitem{ze-bo-ha-ma-05b}
N. Zenine, S. Boukraa, S. Hassani, J.M. Maillard,
        J. Phys. A: Math. Gen. {\bf 38} (2005) 4149-4173
 and  arXiv:cond-mat/0502155  

\bibitem{Kawai} M. Kashiwara and T. Kawai, Publ. RIMS, Kyoto Univ. {\bf 12} (Suppl): 131 (1977) 

\bibitem{ze-bo-ha-ma-05c}
N. Zenine, S. Boukraa, S. Hassani, J.M. Maillard,
{\em Square lattice Ising model susceptibility: connection matrices
          and singular behavior of $\chi^{(3)}$ and $\chi^{(4)}$},
          J. Phys. A: Math. Gen. {\bf 38 } (2005) 9439-9474
 and  math-ph/0506065



\bibitem{Smatrix}
R.J. Eden, P.V. Landshoff, D.I. Olive and J.C. Polkinghorne,
\textit{The analytic S-Matrix}, (Cambridge University Press, 1966)

\bibitem{Kreimer} D. Kreimer,  {\em Knots and Feynman Diagrams},
Cambridge Lecture Notes in Physics {\bf 13}, 
Cambridge University Press (2000), Chapter 9. 

\bibitem{landau-59} L.D.Landau, Nucl Phys {\bf 13} (1959) 181

\bibitem{bjor-drel-65}
J.D. Bjorken, S.D. Drell, {\em Relativistic Quantum Fields},
(New York, McGraw-Hill, 1965)

\bibitem{nickel-05} B. Nickel,
 J. Phys. A: Math. Gen. \textbf{38} (2005) 4517-4518


\bibitem{Crandall}
D.H. Bailey, J.M. Borwein and R.E. Crandall,
J. Phys. A {\bf 39} (2006) 12271-12302

\bibitem{Boos2006}
H. E. Boos, F. G\"ohmann, A. Kl\"umper and J. Suzuki,
 {\em The 75th Anniversary of the Bethe Ansatz FREE
Factorization of multiple integrals representing the density matrix of
 a finite segment of the Heisenberg spin chain},
 J. Stat. Mech. (2006) P04001

 
\bibitem{Korepin} H.E. Boos and V.E. Korepin,
J. Phys. A {\bf 34} (2001) 5311-5316

\bibitem{pal-tra-81}
J. Palmer, C. Tracy,
Adv. Appl. Math. {\bf 2} (1981)329

\bibitem{yamada-84}
K. Yamada,
Prog. Theor. Phys. {\bf 71} (1984)1416

\bibitem{itz-zub-80}
C. Itzykson and J.B. Zuber,
{\it Quantum fields},
(McGraw-Hill, New-York, 1980)


\bibitem{lip-poo}
L. Lipshitz, A. J. van der Poorten,  {\it Rational functions, 
diagonals, automata and arithmetic}, Proc. First Conf. of the Canadian Number
Theory Association, in Number theory (Banff, AB, 1988), editor R. A. Mollin, 
 pages 339-358. de Gruyter, Berlin, 1990.

\bibitem{Hadamard}
J. Havil, {\em Gamma: Exploring Euler's constant},
(Princeton University Press, Princeton NJ, 2003);
A. Voros, Comm. Math. Phys. {\bf 110} (1987) 439-465

\bibitem{christol1}
G. Christol, {\em Diagonales de fractions rationnelles et \'equations 
diff\'erentielles}, Groupe d'\'etude d'analyse ultram\'etrique, (Amice, 
Christol, Robba), Inst. Henri Poincar\'e, Paris, 1982/83 $n^{o}\, 18$


\bibitem{christol2}
G. Christol, {\em Diagonales de fractions rationnelles et \'equations 
de Picard-Fuchs}, Groupe d'\'etude d'analyse ultram\'etrique, (Amice, 
Christol, Robba), Inst. Henri Poincar\'e, Paris, 1984/85 $n^{o}\, 13$


\bibitem{christol3}
A. J. van der Poorten, {\em Power series representing algebraic 
functions}, in S. David ed. S\'eminaire de th\'eorie des nombres,
 Paris 1990-91, Birkh\"auser, Boston, pp. 241-262\\
http://www-centre.ics.mq.edu.au/alfpapers/a103.pdf

\bibitem{necer-97}
A. Necer,
{\it S\'eries formelles et produit d'Hadamard},
J. Th\'eorie des Nombres Bordeaux, {\bf 9} (1997)319-335

\bibitem{fagnot-96}
I. Fagnot,
{\it Langage de Lukasiewicz et diagonales de s\'eries formelles},
J. Th\'eorie des Nombres Bordeaux, {\bf 8} (1996)31-45

\bibitem{allouche-97}
J. P. Allouche,  
{\it On the transcendence of formal power series}, (1997),
Theoret. Comput. Sci. {\bf 218} (1999)143-160

\bibitem{Meijer} J. H. Davenport,
 {\em The Difficulties of Definite Integration}, \\
http://www-calfor.lip6.fr/~rr/Calculemus03/davenport.pdf



\end{thebibliography}
 \end{document}